\documentclass[12pt]{article}
\usepackage{amsmath,amsfonts, amssymb,graphicx,geometry,authblk,color,soul, multirow} 
\DeclareMathOperator{\sign}{sign}

\title{Learning macroscopic internal variables and history dependence from microscopic models}
\author[1]{Burigede Liu\footnote{These authors contributed equally}\footnote{Corresponding authors: \tt{bl377@cam.ac.uk}, \tt{bhatta@caltech.edu}}}
\author[2]{Eric Ocegueda$^{*}$}
\author[2]{Margaret Trautner}
\author[2]{Andrew M. Stuart}
\author[2]{Kaushik Bhattacharya$^{\dagger}$}

\affil[1]{University of Cambridge, Cambridge, United Kingdom}
\affil[2]{California Institute of Technology, Pasadena, United States}
 
\begin{document}
\maketitle

\begin{abstract}
This paper concerns the study of history dependent phenomena in heterogeneous materials in a two-scale setting where the material is specified at a fine microscopic scale of heterogeneities that is much smaller than the coarse macroscopic scale of application.  We specifically study a polycrystalline medium where each grain is governed by crystal plasticity while the solid is subjected to macroscopic dynamic loads.  The theory of homogenization allows us to solve the macroscale problem directly with a constitutive relation that is defined implicitly by the solution of the microscale problem.   However, the homogenization leads to a highly complex history dependence at the macroscale, one that can be quite different from that at the microscale.  In this paper, we examine the use of machine-learning, and especially deep neural networks, to harness data generated by repeatedly solving the finer scale model to: (i) gain insights into the history dependence and the macroscopic internal variables that govern the overall response; and (ii) to create a computationally efficient surrogate of its solution operator, that can directly be used at the coarser scale with no further modeling.  We do so by introducing a recurrent neural operator (RNO), and show that: (i) the architecture and the learned { internal} variables can provide insight into the physics of the macroscopic problem; and (ii) that the RNO can provide multiscale, specifically FE$^2$, accuracy at a cost comparable to a conventional empirical constitutive relation.

\end{abstract}

\section{Introduction}

The engineering properties of materials are determined through complex interactions of mechanisms that operate over a range of time and length scales (e.g. \cite{phillips2001crystals}).  Over the last few decades, the framework of multiscale modeling has emerged and provides means of understanding this complexity.   The entire range of material behavior is divided into a hierarchy of scales, the relevant phenomenon is identified and modeled at each scale and the hierarchy is reassembled assuming a pair-wise interaction between the scales (e.g., \cite{van2020roadmap}).  Statistical mechanics and the theory of homogenization provide the basis for this pair-wise reassembly: the coarser scale controls or regulates the finer scale, and observes the averaged or filtered response from the smaller scale.

Of particular interest in this work is the link between two continuum scales where the physics is specified on a scale that is significantly smaller than that of the application.  Consider for example a polycrystalline solid with grains significantly smaller than the overall dimensions where each grain is governed by crystal plasticity while the solid is subjected to macroscopic loads.  The theory of homogenization \cite{bensoussan2011asymptotic,pavliotis2008multiscale}  states that we may solve the balance laws at the application or macroscopic or coarse scale using a constitutive relation that is determined in an implicit manner by solving a problem on a unit cell or representative volume at the microscopic or fine scale at which the physics is specified. The implementation of this framework is prohibitively computationally expensive since one needs to solve a microscopic scale unit cell problem at each point and at each time of the macroscopic calculation.  This provides the first motivation of our work.

Also of interest in this work is history and rate-dependent phenomena (like crystal plasticity with hardening and rate dependence). While the history dependence is specified at the microscopic scale, the process of homogenization or averaging over the unit cell adds significant complexity. In particular, heterogeneous strain and strain-rate distributions within the unit cell mean that phenomena at distinct rates interact and thus the history dependence at the macroscopic scale can be very different. For example, it is easy to show that a composite material made of constituents governed by a linear Kelvin-Voigt model where the stress depends only on strain and strain-rate can have infinite memory (i.e., the stress depends on the entire history of strain) at the macroscopic scale (e.g., \cite{sanchez_homogenization_1978,francfort1986homogenization}). In fact, Brenner and Suquet \cite{brenner2013overall} provide examples where the macroscopic response can only be described using a fractional time derivative of the strain. The situation in nonlinear history dependent phenomena like plasticity is far more complex and very little is known theoretically.  This provides the second motivation for our work.

These issues have motivated a growing body of work in computational mechanics that seeks to use machine learning in multiscale modeling.  { In an early work, Lefik, Boso and Shrefler \cite{lefik_2009} apply machine learning to the multiscale modeling of composite materials.   More recently,} various researchers have used a recurrent neural network (RNN) to approximate history dependent behavior.  Ghavamian and Simone \cite{ghavamian_accelerating_2019} suggested that an RNN with long short-term memory (LSTM) can be used to describe effective plastic behavior of a representative volume and then be used in FE$^2$ calculations.  They demonstrated it in an ``academic example'' of a bar governed by isotropic Perzyna viscoplasticity with periodic line of holes.    Mozaffar {\it et al.} \cite{mozaffar2019deep} showed that an RNN with gated recurrent units (GRUs) is able to approximate the homogenized behavior of a composite medium consisting of isotropic elastic inclusions in an isotropic rate-independent plastic matrix.  Wu {\it et al.} \cite{wu_recurrent_2020} used an RNN to approximate the overall unit cell behavior of a particulate composite consisting of dilute circular elastic particles in an isotropic matrix undergoing  J2 plasticity,  isotropic hardening and no rate-effects, and then used the trained macroscopic (GRU) RNN in macroscopic simulations.   Wu and Noels \cite{wu_recurrent_2022} seek to  predict macroscopic response and microscopic strain distribution using a (GRU) RNN combined with PCA of the strain-field. They again apply it to a particulate composite consisting of dilute circular elastic particles in an isotropic matrix undergoing J2 plasticity, isotropic  hardening and no rate-effects.  { There is a similar study of plasticity in particulate composites by  Logarzo, Capuano and Rimoli \cite{logarzo_2021}.}  There is also an emerging literature in using RNNs to accelerate large scale finite element calculations (e.g. \cite{im_surrogate_2021,bonatti_recurrent_2022}).  One issue in this approach is that the complexity of the LSTM/GRU architecture with very large number of parameters (e.g., millions in \cite{mozaffar2019deep}).

In this paper, we examine the use of machine-learning and especially deep neural networks to harness data generated by repeatedly solving the finer scale model: (i) to gain insights into the history dependence and the macroscopic internal variables that govern the overall response; and (ii) to create a computationally efficient surrogate of its solution operator, that can directly be used at the coarser scale with no further modeling.

We draw inspiration from the established literature on the use of internal or state variable theories in continuum mechanics \cite{rice1971inelastic,gurtin_cont_mech} (and more broadly the use of order parameters in physics) to develop the architecture for our machine-learnt approximation.  These theories postulate that all prior history of deformation can be represented through state or internal variables that evolve with the deformation.  So the stress at any point at any instant of time is determined locally from the deformation gradient and the state variable at that point at that instant using a ``stress-strain'' relation.  Further, the rate of change of the state variable at any point at any instant of time again depends locally on the deformation gradient and state variable at that point at that instant through a ``kinetic relation'' or ``evolution law''.   Therefore, we introduce two feed-forward neural networks, one for the stress-strain relation and one for the kinetic relation.  Importantly, { the internal variables are not specified {\it a priori}, but are discovered through training using macroscopic data.}   

Further, we formulate the approximation in continuous time and therefore this architecture represents a functional that maps the history of the deformation gradient to the stress.    In practice, both the training data and the application involve discretized data and so one could formulate the architecture for the discrete data.  However, in such situations, the architecture and the training embeds the discretization, and it provides poor approximation when applied to other discretization. This is problematic because the training and application of the approximation may involve different time discretizations.  This is indeed the case in multiscale modeling since the coarse scale time-step is dictated by the application and the coarse-scale numerical method.  We overcome this by formulating the approximation in a time-continuous manner, and then discretizing as necessary during training and application.  { An alternate approach in the context of LSTM-based RNNs has recently been proposed by Bonatti and Mohr \cite{bonatti2022}.}

We describe the resulting architecture as a recurrent neural operator (RNO).   We use the same feed-forward networks at every instant of time and therefore, we regard this architecture as ``recurrent''.    Further, we use the word ``operator'' to emphasize the time continuous formulation.  Once discretized in time, the resulting architecture is similar to an RNN.  However, it differs from the usual standard LSTM or gated RNNs in the way it introduces history.  In fact, our architecture is much simpler, and requires a significantly smaller number of parameters, and thus less training data.

 Here, we focus on elasto-viscoplastic polycrystals.   The behavior of the single crystal is governed by crystal plasticity and therefore can be highly anisotropic. After some lower dimensional examples, we study the case of magnesium where we have a combination of soft slip systems, intermediate tension twin systems and hard slip systems. Further, we have isotropic rate-dependent hardening of these materials. Therefore, we expect the macroscopic behavior to be complex with non-trivial history dependence.  We use repeated solutions of the unit cell problem subject to various average deformation gradient trajectories to generate the data in the form of average deformation gradient trajectory, average stress pairs.  We use this data to train our approximation and show that the  proposed RNO provides an accurate and efficient representation.   We examine the insights that the  RNO can  provide into hidden physics from data about the deformation gradient and stress.   As already noted, we anticipate  highly complex history dependence in materials, and there is no established method to identify the macroscopic internal variable (or even its dimensionality) either through homogenization theory or experimentally.  So even identifying the number of  internal variables required to represent the data can be meaningful.  In some cases, it is also possible to obtain further insight.  We then demonstrate that the trained RNO can be used effectively and efficiently as a (macroscopic) constitutive relation in macroscopic calculations.   Crucially, we show convergence with respect to time-discretization, and find that the cost of using the RNO in a macroscopic calculation is only marginally more than that of a classical constitutive relation.  In other words, the trained RNO provides provides multiscale, specifically FE$^2$, accuracy at a cost comparable to a conventional empirical constitutive relation.
 
The paper is organized as follows.  Section \ref{sec:arch} provides the background and describes the proposed architecture and training procedure.  We then study an academic example of a one-dimensional composite in Section \ref{sec:1d}.  We can explicitly compute the effective response in this example, and we show that the proposed architecture and training lead to an accurate approximation.  { We study a second academic example of three dimensional laminates in Section \ref{sec:lam}.  We show that the RNO and training leads to an accurate approximation, but with fewer internal variables than that suggested by theory.  We explore this with a detailed study of the RNO trained with macroscopic response and the micromechanical internal variables.}  We turn to polycrystals in Section \ref{sec:learn}.  We first study a two-dimensional example where each grain has two slip systems and use data generated by full-field numerical simulations.  We show that the RNO is able to provide a highly accurate surrogate, and that the { internal} variables can provide some insights into the underlying physics.  We subsequently turn to the case of magnesium in three dimensions with data generated by Taylor averaging. Again, we show that an RNO is able to provide an accurate surrogate.  We discuss the link between resolution independence and architecture in Section \ref{sec:res} and  demonstrate resolution independence of the proposed RNO.  We then use the trained  RNO for magnesium as a surrogate in macroscopic problems in Section \ref{sec:multi}.  We conclude in Section \ref{sec:disc} with a discussion.

\section{Background, proposed architecture and training} \label{sec:arch} 

We are interested in problems where there is a separation of scales between the scale at which the physics is specified (e.g., the crystal plasticity model at the scale of an individual grain), and the scale of the application (e.g., an impact problem).  The theory of homogenization \cite{bensoussan2011asymptotic,pavliotis2008multiscale} states that we may solve the balance laws at the application scale using a constitutive relation 
\begin{equation} \label{eq:const}
\Psi: \{ { F}(\tau): \tau \in (0,t) \} \mapsto \sigma (t)
\end{equation}
that links the macroscopic { deformation gradient history} to the macroscopic Cauchy stress, and one that is determined in an implicit manner by solving the fine scale problem at which the physics is specified on a unit cell or representative volume.   The { macroscopic deformation gradient} history provides the boundary condition for the unit cell problem and the returned macroscopic stress is the average of the stress in the unit cell.  We refer the reader to \cite{liu2022learning} for details.

This framework raises two issues.  The first concerns implementation of the framework: the evaluation of this map $\Psi$ is prohibitively expensive in most problems of interest since the unit cell problem has to be solved at each instant of time (numerically at each time step) at each material point (numerically at each quadrature point).  Therefore, it would be useful to have an approximation that is both accurate and inexpensive to evaluate.  The second concerns an understanding of the underlying physics.  Since the unit cell problem involves the solution of a complex system of equation, the nature of the dependence of the stress on strain history at the macroscopic scale, and whether such history can be encoded into a { internal} variable is unclear.

We address these two issues by seeking to approximate the map $\Psi$ using a recurrent neural operator (RNO) in this work.   Specifically, we consider an RNO approximation of the form 
\begin{equation} \label{eq:RNO}
\begin{cases} 
\sigma(t) = f({F}(t), \{\xi_\alpha (t) \}_{\alpha=1}^k), \\
\dot{\xi}_i(t) = g_i ({F}(t), \{\xi_\alpha(t) \}_{\alpha=1}^k) \quad i = 1, \dots k 
\end{cases}
\end{equation}
where $f: { {\mathbb R}^{d\times d}} \times {\mathbb R}^k \to {\mathbb R}^{d \times d}_\text{symm}$ and $g_i: { {\mathbb R}^{d\times d}} \times {\mathbb R}^k \to {\mathbb R}$ are neural networks and the superposed dot represents time derivative.  We call $\{\xi_\alpha(t) \}_{\alpha=1}^k$ the {\it { internal} variables}, and these encapsulate the history.  We do not identify the { internal} variables {\it a priori}; instead, they are identified as a part of the training process. Note that while the architecture maps trajectories to trajectories, the architecture is local and identifies the stress simply on the value of the deformation gradient and stretch at that instance.

We now make a series of comments about the RNO and its training, as well as a few other alternatives.

\paragraph{Discretization and discretization independence}

The RNO (\ref{eq:RNO}) is an operator that maps functions ({ deformation gradient history} over a time interval) to a matrix (stress).  However, in practice, both training and application data are discretized in time, although they do not necessarily have the same discretization/resolution. In order to construct an architecture that is independent of the time discretization, we use the following forward Euler discretization: for a time-discretization with time-steps $\{(\Delta t)_n\}$, we evaluate the RNO  at the $n^\text{th}$ time step to be
\begin{equation} \label{eq:RNOd}
\begin{cases} 
\sigma^n = f({F}^n, \{\xi_\alpha^{n} \}_{\alpha=1}^k), \\
\xi^n_i =  \xi^{n-1}_i + (\Delta t)_n \ g_i ({ F}^n, \{\xi_\alpha^{n-1} \}_{\alpha=1}^k) \quad i = 1, \dots k .
\end{cases}
\end{equation}
In this work, we use uniform discretization so that $(\Delta t)_n = (\Delta t)$ is independent of $n$.  
For this approximation to be meaningful, it should be independent of the time discretization.  In other words, we should be able to train it at one time discretization and use it at another time discretization without any loss of accuracy.  This is also important in practice because the training data may need to be generated at smaller time steps due to the underlying microscale physics compared to that used in a macroscale application.

{
\paragraph{RNO architecture} 
We use fully connected neural networks \cite{lecun2015deep} to model the feed-forward functions $f$ and $g$. By default, we assume that each of $f$ and $g$ is comprised of four intermediate linear layers, each with 100 nodes. Nonlinear activation functions are applied using the scaled exponential linear unit (SELU) \cite{klambauer2017self}, and the networks are trained with the ADAM optimization algorithm \cite{kingma2014adam}.

}

\paragraph{Training and training data}
We train the RNO, i.e., identify the parameters of the neural networks $f, g_i$ as well as the { internal} variables $\{\xi_\alpha\}_{\alpha=1}^k$, using data generated by the numerical solution of the unit cell problem.  Specifically, we repeatedly solve the unit cell problem to obtain various realizations of the map 
\begin{equation} \label{eq:map}
\Psi_T: \{ { F} (t): \tau \in (0,T) \} \mapsto \{  \sigma  (t): t \in (0,T)\}.
\end{equation}
for some fixed $T>0$.

A crucial issue in generating numerical data is to balance the cost of generating the data with the need to sample sufficient input (average { deformation gradient} histories) to provide an accurate enough approximation for the input encountered in application. This leads to the question of identifying an optimal distribution of input.  This remains an active area of research.  In our application on impact problems, we anticipate encountering trajectories of macroscopic stretch that vary smoothly with time, but change directions arbitrarily.   To this end, we divide $(0,T)$ into $M$ intervals $(\Delta t)^m = t^m - t^{m-1}, m = 1, \dots M$ where $0 = t^0 < t^1 < \dots < t^M=T$ and set 
{ $(F_{ij})^m = (F_{ij})^{m-1} + (\nu_{ij})^m F_\text{max}\sqrt{(\Delta t)^m},$ $i,j = 1, \dots, d, i \le j
$
where $(\nu_{ij})^m \in \{-1,1\}$ follow a Rademacher distribution for each $ij$}.  We take $F_{ij}(t)$ to be the cubic  Hermite interpolation of $\{ (t^m, (F_{ij})^m) \}$.  We take $T=200$, $M=10$ and random time intervals drawn from a uniform distribution.    We clarify that the subintervals $(\Delta t)^m$ used here to determine loading paths are distinct from the time step $(\Delta t)$ that we use to either train or use the RNO.

 
\paragraph{{ Internal} variables}
 In our formulation, we have to decide {\it a priori} the number $k$ of { internal} variables (though not the { internal} variable themselves).  Our approach is to train the neural network separately for various numbers of { internal} variables, and then test the resulting approximation over a set of test data.  We find that the test error initially decreases with increasing $k$, but then saturates beyond a particular value.  In fact, we find the transition from decrease to saturation is sharp.  We take the value of $k$ at this sharp transition to be the number of { internal} variables.
 
 As already noted the { internal} variables have no inherent meaning. In fact, it is easily verified from (\ref{eq:RNO}) that any reparametrization or smooth one-to-one and onto change of variables $\{\xi_\alpha\} \to \{\eta_\alpha\} $ of the form
\begin{equation} \label{eq:cov}
\eta_i = \varphi_i ( \{ \xi_\alpha\} ) 
\end{equation}
yields an equivalent RNO:
\begin{equation}
\begin{cases} 
\sigma(t) = \hat{f}({ F}(t), \{\eta_\alpha (t) \}_{\alpha=1}^k), \\
\dot{\eta}_i(t) = \hat{g}_i ({F}(t), \{\eta_\alpha(t) \}_{\alpha=1}^k) \quad i = 1, \dots k 
\end{cases}
\end{equation}
where 
\begin{align*}
\hat{f}({F}, \{\eta_\alpha\})_{\alpha=1}^k) &= f(({F}, \{\varphi^{-1}_\alpha (\{\eta_\beta\}_{\beta=1}^k ) \})_{\alpha=1}^k), \\
\hat{g}_i ({F}, \{\eta_\alpha\})_{\alpha=1}^k) &= \sum_{j=1}^k \frac{\partial \varphi_i}{\partial \xi_j}  (\{\eta_\alpha\}_{\alpha=1}^k) 
g_j (({F}, \{\varphi^{-1}_\alpha (\{\eta_\beta\}_{\beta=1}^k ) \})_{\alpha=1}^k).
\end{align*}

This invariance also has implications on the training: the same data can produce different (but equivalent) choices of the { internal} variable depending on the initialization.

\paragraph{Internal or state variable theories}
The form of the RNO (\ref{eq:RNO}) is similar to the internal or state variable theories that are widely used to describe inelastic behavior in continuum mechanics and physics \cite{rice1971inelastic,gurtin_cont_mech}.  We may regard the { learned internal variables those} variables that represent the state of the material; we take the first of the two equations to be the stress-strain relation and the second to be the kinetic relation for the evolution of the state variables.  However, the choice of  state variables postulated in theories of mechanics reflect physical intuition, while those in the RNO reflects the behavior of the training data and is not unique due to the invariance under reparametrization (\ref{eq:cov}).

\paragraph{ Thermodynamic and symmetry restrictions}{ In this work, we do not impose any thermodynamic or material symmetry restrictions on the functions $f, g_i$ in (\ref{eq:RNO}).  Physics as well as mathematical well-posedness require the constitutive relations to satisfy some  thermodynamic restrictions.  In our setting, we expect the stress $f$ to be the derivative of a quasiconvex strain energy functional while we expect the evolution law $g_i$ to be dissipative (or the derivative of a non-negative and convex dissipation potential).  Similarly, in some problems we may know {\it a priori} the overall material symmetry.  For example, in the polycrystals we study in Section \ref{sec:learn}, we expect the overall behavior to be isotropic.  It is possible to build architectures with these conditions (e.g., \cite{masi_2021} for dissipative processes and \cite{as'ad_2022} for hyperelasticity).  We do not do so.  However, since our data is generated from numerical solutions of well-posed problems, we anticipate that the model automatically learn such restrictions from the data.  For example, we demonstrate in \cite{liu2022learning} that a PCA-Net automatically learns isotropy from the data.   We discuss this further in Section \ref{sec:disc}.
}

\paragraph{Viscoelastic RNO}
In the RNO in (\ref{eq:RNO}), we do not include an explicit dependence on the rate of change of deformation gradient $\dot F$.   So we can introduce an alternate form
\begin{equation} \label{eq:RNO2}
\begin{cases} 
\sigma(t) = \bar{f}({F}(t), \dot{{F}}(t), \{\xi_\alpha (t) \}_{\alpha=1}^k), \\
\dot{\xi}_i(t) = \bar{g}_i ({F}(t), \{\xi_\alpha(t) \}_{\alpha=1}^k) \quad i = 1, \dots k 
\end{cases}
\end{equation}
that explicitly includes the rate of change of $\dot F$.  We call this the {\it viscoelastic RNO}.

The viscoelastic RNO (\ref{eq:RNO2}) includes the original form (\ref{eq:RNO}) as a special case.  However, since $\bar{f}, \bar{g_i}$ are over-parametrized neural networks, it would take a very large amount of data for the training to recognize it.  Therefore, they tend to behave differently, especially with respect to discretization independence.  We shall show in what follows that for elasto-viscoplasticity that we study in the present paper, the form (\ref{eq:RNO}) displays discretization independence while (\ref{eq:RNO2}) does not.  In contrast, we have studied the same issues in { Kelvin-Voigt} viscoelasticity elsewhere \cite{viscoelasticity}, and we observe there that the form (\ref{eq:RNO2}) displays discretization independence while (\ref{eq:RNO}) does not as one would expect from the differing physics.

\paragraph{PCA-Net}  In our previous work \cite{liu2022learning}, we studied 
use a different neural operator architecture, named PCA-Net, to approximate the map (\ref{eq:map}). This architecture takes the stretch history over the entire interval $[0,T]$ to the stress history over the entire interval $[0,T]$ and,
therefore, is not causal: the stress at any instance $t$ depends on the stretch over the interval both before and after $t$.  We found that the approximation learnt causality from the data, however. In contrast, causality is built into the proposed RNO (\ref{eq:RNO}).  Similarly, that architecture is not local, but the current one is.

\section{Elasto-viscoplastic composites in one dimension} \label{sec:1d}

In this, and the following section, we study the ability of the proposed RNO to learn the map $\Psi$ at the macroscopic scale using data from the repeated solution to a unit cell problem.  We formulate a crystal plasticity model at the microscopic scale, solve these equations repeatedly over a unit cell subjected to a various (average) stretch trajectories as the boundary condition, and calculate the corresponding stress trajectory.   We then use the resulting pairs of average stretch and average stress trajectories to train the RNO.

We begin with a simple example in one dimension.  It is customary to work with displacement and strain $\varepsilon$, and we do so noting that $\varepsilon = {F}-1$ in the notation introduced earlier.  We begin by formulating the unit cell problem. The governing equations are
\begin{align}
&\varepsilon = \frac{\partial u}{\partial x}, \quad
\sigma = E (\varepsilon - \varepsilon_p), \quad
\frac{d\sigma}{dx} = 0, \quad
\dot{\varepsilon}_p = \dot{\varepsilon}_{p0} \sign (\sigma) \left(\frac{|\sigma|}{\sigma_0}\right)^n, \label{eq:1dfield}\\
& u(x,0) = 0, \quad \varepsilon_p(x,0) = 0, \label{eq:1dic}\\
& u(0,t)= 0, \quad u(1,t) =  \bar{\varepsilon}(t) \label{eq:1dbc}.
\end{align}
where the displacement $u$, the strain $\varepsilon$ , the plastic strain $\varepsilon_p$, the stress $\sigma$ are the variables, and the material parameters are Young's modulus $E$, rate constant\footnote{It is customary to use a superposed dot to denote this material constant $\dot{\varepsilon}_{p0}$; note that it does not denote time derivative in this context.} $\dot{\varepsilon}_{p0}$, yield stress $\sigma_0$ and rate exponent $n$. The first line of equations are the field equations (compatibility, stress-strain, equilibrium and plastic flow rule in order), the second line the initial conditions and the final line are the boundary conditions. We consider a composite material made of $N$ constituents with material parameters $E$, $\dot{\varepsilon}_{p0}$, $\sigma_0$ and $n$ are piecewise constant.

We can rewrite the constitutive relation (\ref{eq:1dfield})$_2$ as $\varepsilon - \varepsilon_p = E^{-1} \sigma$.  We observe from compatibility (\ref{eq:1dfield})$_1$ and boundary conditions (\ref{eq:1dbc}) that $\langle \varepsilon \rangle = \bar \varepsilon$ where $\langle \cdot \rangle$ denotes average over $x$, and from equilibrium (\ref{eq:1dfield})$_3$ that $\sigma$ is independent of $x$.  Therefore $\bar{\varepsilon} - \langle \varepsilon_p \rangle = \langle E^{-1} \rangle \sigma$.  We conclude that 
\begin{align}
&\sigma = \langle E^{-1} \rangle^{-1} ( \bar{\varepsilon} - \langle \varepsilon_p \rangle),  \\
&\langle \dot{\varepsilon}_p \rangle =  \langle \dot{\varepsilon}_{p0} \ \sigma_0^{-n} \sign (\sigma) |\sigma|^n \rangle 
= \sign  ( \bar{\varepsilon} - \langle \varepsilon_p \rangle)  \left\langle \dot{\varepsilon}_{p0} \ \sigma_0^{-n}    \langle E^{-1} \rangle^{-n}  | \bar{\varepsilon} - \langle \varepsilon_p \rangle|^{n} \right\rangle .
\end{align}
Setting $\xi =  \langle \varepsilon_p \rangle$, we find that the effective behavior is of the form (\ref{eq:RNO}) with one internal variable,
\begin{align} \label{eq:1d}
\begin{cases}
\sigma = f(\bar{\varepsilon}, \xi),\\
\dot{\xi} = g (\bar{\varepsilon}, \xi). 
\end{cases}
\end{align}
Therefore, we conclude that we should be able to represent the effective behavior with a single internal variable.  We further note that in this particular case, $f,g$ are in fact functions of a single variable, $\eta = \bar{\varepsilon} - \xi$:
\begin{align}
&\sigma = f(\bar{\varepsilon}, \xi) = \tilde f(\bar{\varepsilon} - \xi), \label{eq:ftilde}\\
&\dot{\xi} = g (\bar{\varepsilon}, \xi) = \tilde g (\bar{\varepsilon} - \xi). \label{eq:gtilde}
\end{align}

We now examine, in a computational example, whether we are able to recover this structure from simply the strain history to stress map. We consider composite materials with $N$ pieces of equal length, with $N$ ranging from 1 to 10.  For each $N$,  we sample $E$, $n$ and $\sigma_0$ i.i.d. from a uniform distribution supported on $[1,10] \text{GPa}$, $[1,20]$ and $[0.1, 1] \text{GPa} $ respectively.  
In all cases $\dot{\epsilon}_{p0}$ is fixed to be $ 1 s^{-1}$.


We solve the governing equations (\ref{eq:1dfield}), (\ref{eq:1dic}) and (\ref{eq:1dbc}) above to generate the data in the form of a homogenized strain history to stress history map: $\{\bar{\varepsilon}(t), t\in (0,T) \} \mapsto \{\sigma(t), t\in (0,T) \}$.  Note that the equilibrium condition requires the stress to be uniform and independent of $x$.   We sample the input $\{\bar{\varepsilon}(t), t\in (0,T) \}$ along smooth but increasing and decreasing trajectories following the methodology described above.  

We then use this data to train an RNO with the architecture described in Section \ref{sec:arch} with a fixed number of internal variables.  For each $N$, we use 300 samples to train and the remaining 100 to test the RNO. 

%

\begin{figure}[t]
\includegraphics[width=6.5in]{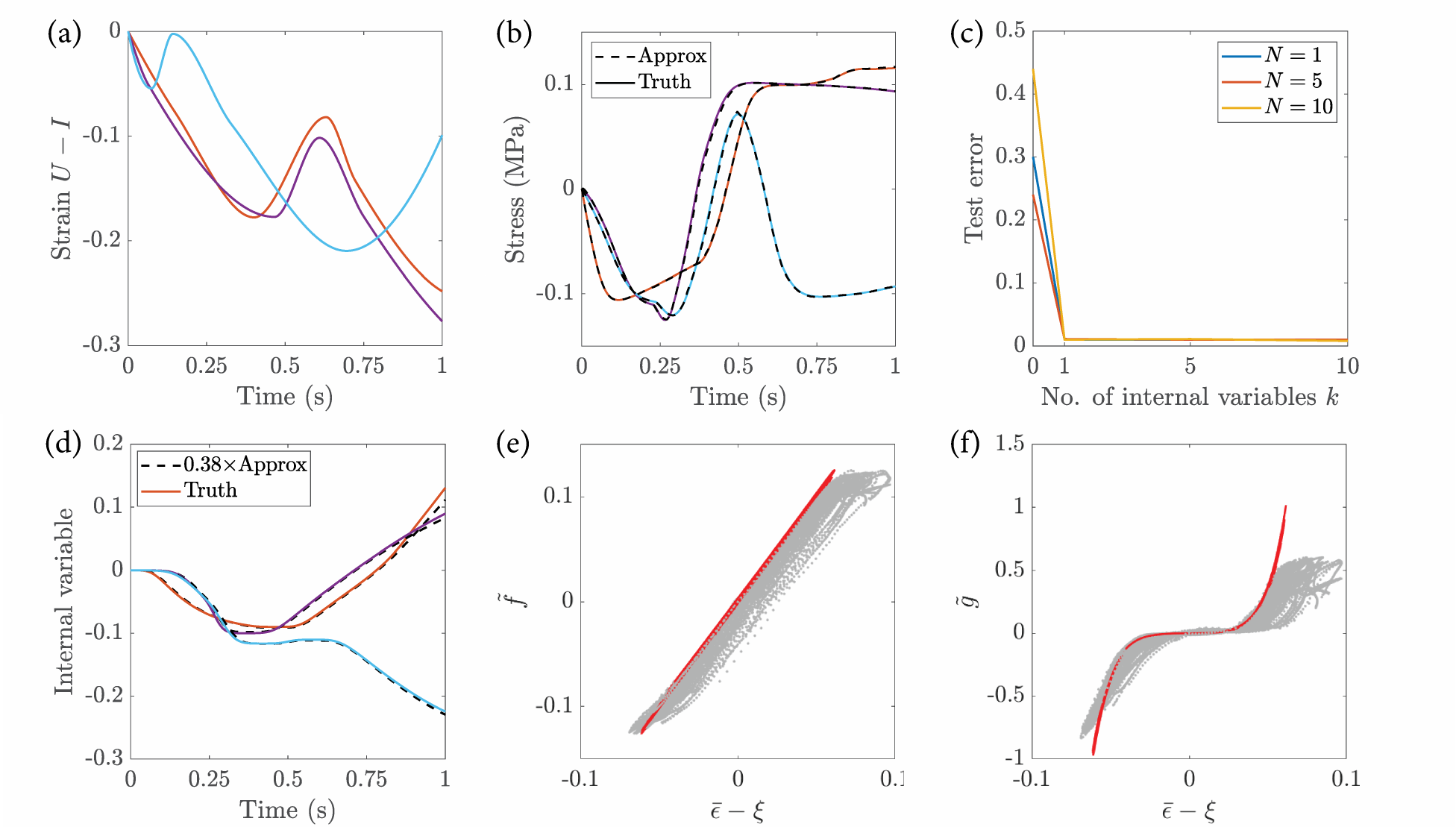}
\caption{Approximating elasto-viscoplasticity in one dimension with a recurrent neural operator.  
(a) Three typical inputs used to train and test the RNO.  
(b) Comparison of the result of the unit cell problem of a three-phase material ($N=3$) corresponding to the inputs in (a) (``Truth") with the output of the trained RNO with one { internal} variable ($k=1$) with the same input (``Approx'').  
(c) The average test error for varying number of internal variables, and for varying number of constituents. 
(d) A comparison of the learnt and actual internal variable for a typical instance of test data in the case of a homogenous material.
(e) Comparison of the actual and learnt function $\tilde f$.
(f) Comparison of the actual and learnt function $\tilde g$. 
}
\label{fig:1d}
\end{figure}
Figure \ref{fig:1d} shows our results.   A typical input is shown in Figure \ref{fig:1d}(a).  The corresponding result of the unit cell problem of a three-phase material ($N=3$) as well as the outputs of the trained RNO with one { internal} variable ($k=1$) are compared in Figure \ref{fig:1d}(b).  We see that the trained RNO is able to accurately capture the actual behavior.  This is studied in detail in Figure \ref{fig:1d}(c) 
which shows the average test error, defined as 
\begin{equation}\label{eq:error}
\text{error} = \frac{1}{S} \sum_{s=1}^S \left(\frac{\int_0^T |\sigma_s^\text{truth}(t) - \sigma_s^\text{approx}(t)|^2 dt}{\int_0^T |\sigma_s^\text{truth}(t)|^2 dt}\right)^{1/2}
\end{equation}
where $s$ indexes the $S$ inputs, for a number of different experiments with varying number of constituents and varying number of internal variables.  We observe that an RNO with one  internal variable is able to learn the overall behavior with high accuracy.  An RNO with no internal variables leads to large errors and increasing the number of internal variables beyond one does not lead to any improvements.   Figure \ref{fig:1d}(d) compares the learnt internal variable (with suitable scaling) with the theoretical internal variable $\langle \varepsilon_p \rangle$.  As noted above, our architecture is invariant under an invertible change of variables; in this case, we find that the training results in a { internal} variable that is simply a scaled version of the original variable.   We observe that our architecture is able to learn the physically meaningful internal variable from the strain history to stress history map.

Finally we examine the efficacy of our training to identify the correct forms for $\tilde f, \tilde g$.  Figures \ref{fig:1d}(e,f) compare the actual functions $\tilde f, \tilde g$ according to (\ref{eq:ftilde}, \ref{eq:gtilde}) with those that are learnt.  The figures shows in dark red the actual functions according to the model, and in light grey dots the points  $\{ \bar{\varepsilon} - \xi, \tilde{f} (\bar{\varepsilon} - \xi) \} $ and $\{ \bar{\varepsilon} - \xi, \tilde{g} (\bar{\varepsilon} - \xi) \}$ for the learnt $\tilde f, \tilde g$ for input strain paths in Figure \ref{fig:1d}(a).   To do this comparison we use the scaled { internal} variable using the scaling identified in Figure \ref{fig:1d}(d).  We see that the training  identifies the two functions to a reasonable degree of accuracy.  

In summary, in the case of a one-dimensional elasto-viscoplastic composite, where we can analytically characterize the effective behavior, we find that our proposed RNO architecture and training procedure is able to accurately learn the effective behavior.

{
\section{Elasto-viscoplastic laminated composite} \label{sec:lam}

We now study a laminated composite where we have alternating layers of equal width of two different elasto-viscoplastic materials at the microscopic scale.  We postulate a plasticity model in infinitesimal strain at the microscopic scale, solve these equations repeatedly over a unit cell subjected to various (average) strain trajectories as the boundary condition, and calculate the corresponding stress trajectory.   We then use the resulting pairs of average strain and average stress trajectories to train the RNO.

\begin{figure}
	\centering%
    \includegraphics[width=4in]{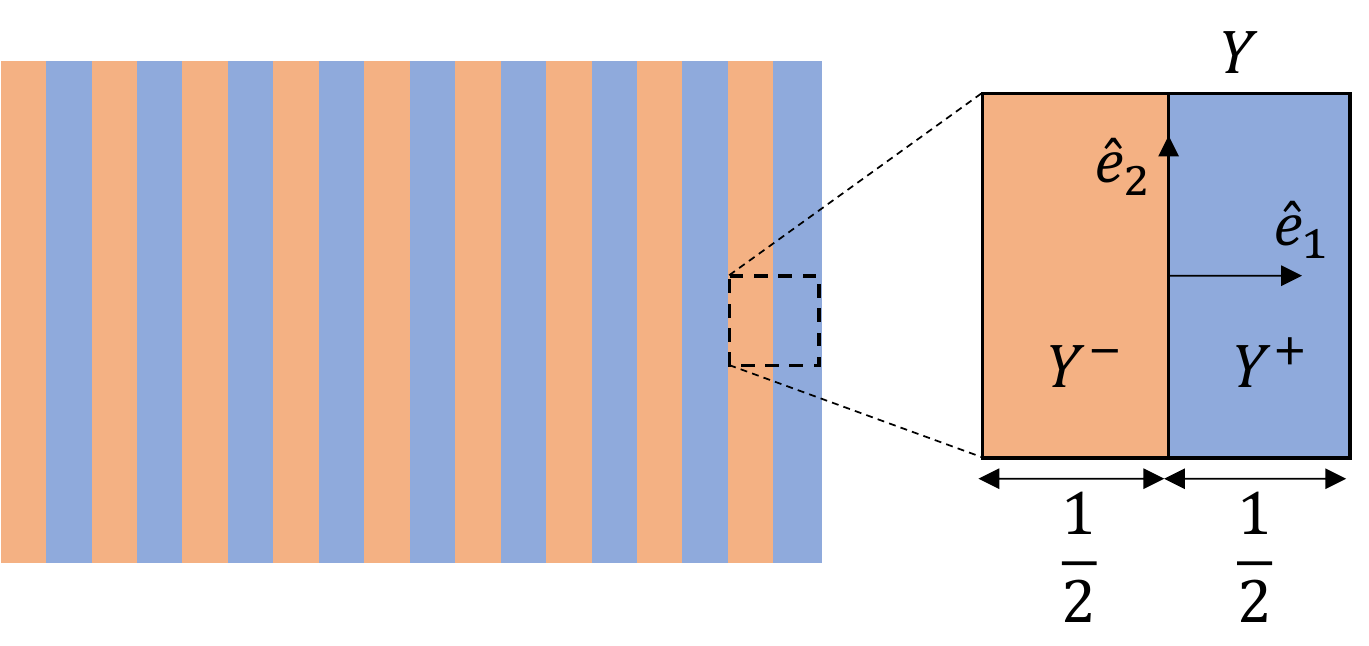}\\
	\caption{A laminate and the unit cell}
	\label{fig:lam}
\end{figure}

The governing equations in the unit cell are 
\begin{eqnarray}
&&\varepsilon = \frac{1}{2}( \nabla u + \nabla u^T), \quad 
\nabla \cdot \sigma = 0,   \label{eq:lamge1}\\
&&\sigma =  {\mathbb C} (\varepsilon - \varepsilon^p), \quad
\dot{\varepsilon}^p = \frac{3}{2}\frac{s}{\bar{s}} \dot{q},  \label{eq:lamge2}\\
&&-\bar{s} + \sigma_0\left(H(q) + \left(\frac{\dot{q}}{\dot{q}_0}\right)^m \right) \ge 0, \quad \dot{q} \ge 0, \quad
\left(-\bar{s} + \sigma_0\left(H(q) + \left(\frac{\dot{q}}{\dot{q}_0}\right)^m \right)\right) \dot{q} = 0  \label{eq:lamge3}
\end{eqnarray}
where $u$ is the displacement, $\varepsilon$ the strain, $\sigma$ the stress, $\mathbb C$ the elastic modulus (assumed to be isotropic with Lam\'e and shear modulus $\lambda$ and $\mu$ respectively, and homogeneous), $\varepsilon^p$ the plastic strain, $q$ the accumulated plastic strain, $s$ the deviatoric stress, $\bar{s} = \sqrt{{\frac32}} |s|$ the effective stress ($|\cdot|$ denoting Frobenius norm), $\sigma_0$ the yield strength, $H(q)$ is the isotropic strain hardening function,
\begin{equation}
H(q) = \begin{cases} 
1  & \text{no strain hardening}\\
1 - \exp \left(- h \frac{q}{\sigma_0} \right) \quad & \text{exponential strain hardening}
\end{cases}
\end{equation}
for some hardening coefficient $h$, $\dot{q}_0$ the reference plastic strain rate and  $m$ the rate-hardening exponent.  The first line describes the strain-displacement and the equilibrium equation, the second line the stress-strain and evolution of the accumulated plastic work and the third line the yield criterion and the flow rule written in terms of a Kuhn-Tucker condition.  We further assume plane stress
\begin{eqnarray}
\sigma_{i3}=0
\end{eqnarray}
and prescribe the history of the in-plane average strain $\bar{\varepsilon}(t)$.  We solve the governing equations subject to the prescribed history of the in-plane average strain $\langle \varepsilon_\text{inplane}(x,t) \rangle_x =  \bar{\varepsilon}(t)$, and find the average in-plane stress $\bar \sigma(t) = $ $\langle \sigma_\text{inplane}(x,t) \rangle_x$.   We are interested in understanding the map
\begin{equation} \label{eq:lammap}
\{ \bar{\varepsilon}(\tau): \tau \in [0,t] \} \mapsto  \bar \sigma  (t).
\end{equation}

For our laminate, the unit cell consists of two layers shown in Figure \ref{fig:lam}. All the
parameters of the model in each of the two layers (such as rate-hardening $m$ for example), are 
distinguished by use of $\pm$ ($m^{\pm}$ for example). We look for a solution to the governing equations (\ref{eq:lamge1}-\ref{eq:lamge3}) that is uniform in the strain, plastic strain, stress and accumulated plastic strain in each region ($\varepsilon^\pm$,$\varepsilon^{p\pm}$, $\sigma^\pm$, $q^\pm$ respectively).  The first two equations (\ref{eq:lamge1}) are automatically satisfied in the interior of each region, and reduce to the jump conditions on the interface
\begin{equation} \label{eq:jump}
\varepsilon^+ - \varepsilon^- = a \otimes \hat{e}_1 + \hat{e}_1 \otimes a, \quad (\sigma^+ - \sigma^-) \hat{e}_1  = 0 .
\end{equation}
We can use this condition along with the plane-stress condition to conclude that
\begin{align}
    a_1 = \frac{\mu}{\lambda + 2\mu} \big( \varepsilon^{\, \text{p} \, +}_{11} - \varepsilon^{\, \text{p} \, -}_{11} \big), \quad
    a_2 = \varepsilon^{\, \text{p} \, +}_{12} - \varepsilon^{\, \text{p} \, -}_{12}, \quad
    a_3 = \varepsilon^{\, \text{p} \, +}_{13} - \varepsilon^{\, \text{p} \, -}_{13} = 0.
\end{align}
Given the average strain history $\bar{\varepsilon}(t)$, we can now solve (\ref{eq:lamge2},\ref{eq:lamge3}) to obtain the average stress history $\bar{\sigma}(t)$ and obtain the map (\ref{eq:lammap}).  Note that we have four { state} variables in each region $( \varepsilon_{11}^{p,\pm}, \varepsilon_{12}^{p,\pm}, \varepsilon_{22}^{p,\pm}, q^\pm )$ for a total of eight { state} variables.

\subsection{No strain hardening} \label{sec:lam_nohard}

\begin{table}
{
\centering
\begin{tabular}{lccccccc}
\hline
Case & $\lambda$ (GPa) & $\mu$ (GPa) & $\sigma_0^+$ (MPa) & $\sigma_0^-$ (MPa) & $m^+$ & $m^-$ & $\dot{q}_0$ (s$^{-1}$)\\
\hline
MH (green) & 24 & 25 & 500 & 500 & 0.2 & 0.2 & 1\\
M1 () & 24 & 25 & 250 & 500 & 0.1 & 0.2 & 1\\
M2 (red) & 24 & 25 & 100 & 500 & 0.04 & 0.2 & 1\\
M3 (yellow) & 24 & 25 & 50 & 500 & 0.02 & 0.2 & 1\\
M4 (purple) & 24 & 25 & 250 & 500 & 0.4 & 0.2 & 1\\
\hline
\end{tabular}
\caption{Material parameters used for laminates \label{tab:lamparam}}
}
\end{table}

\begin{figure}[t!]
    \centering
    \includegraphics[width=6in]{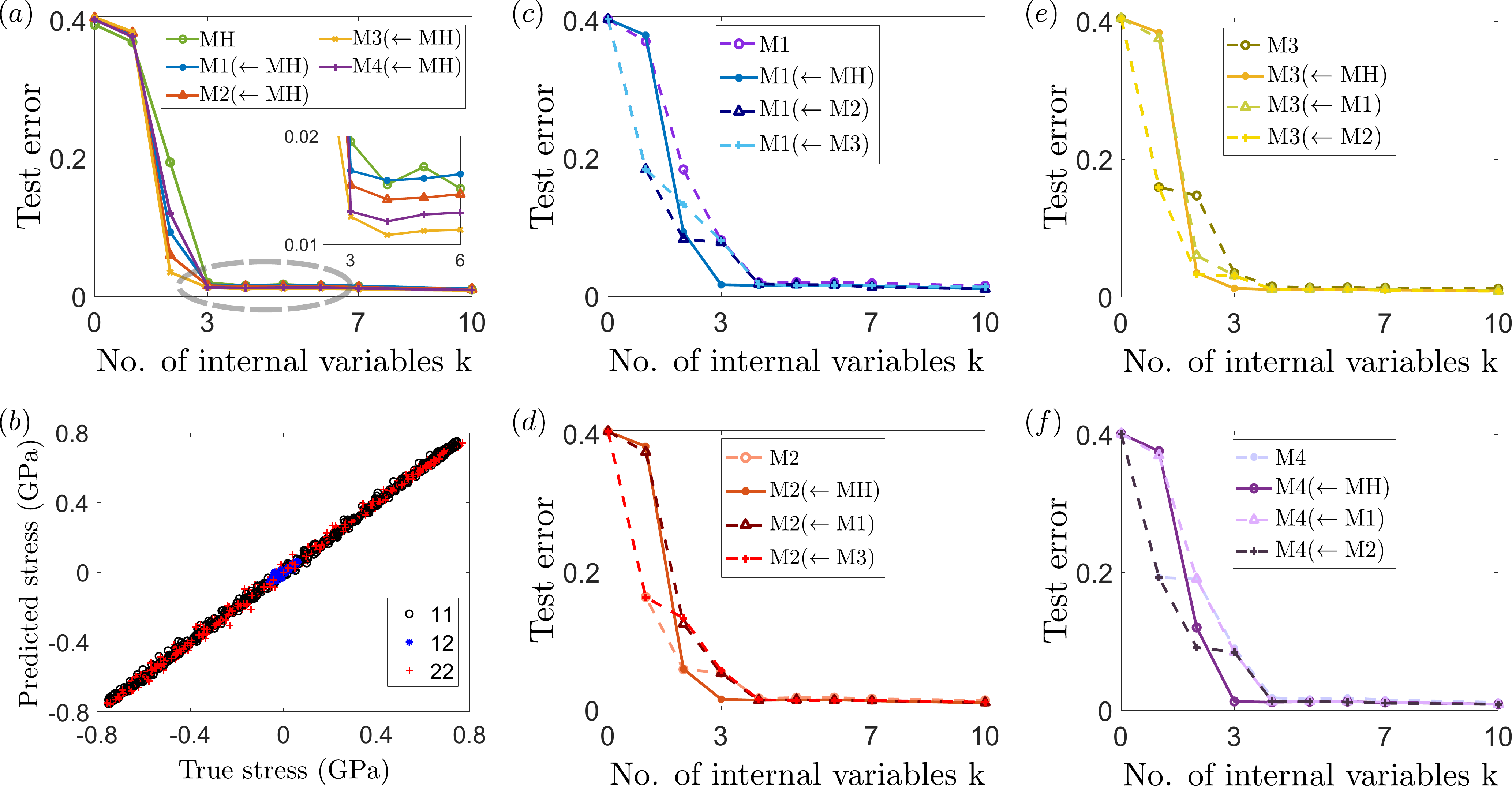}
    \caption{ Approximating a elasto-viscoplastic laminate with no-strain hardening with a recurrent neural operator. (a) Training RNOs with various numbers of { internal} variables.  In each case, we train the homogeneous material (MH) and then use transfer learning.  The notation $A \ (\leftarrow B)$ indicates the case $A$ was trained using transfer learning from $B$.  (b) Comparison of the ground truth and RNO output for eight arbitrarily chosen test strain trajectories for the case M3.  (c-f)  Training RNOs with various numbers of { internal} variables both directly and with transfer learning.
    \label{fig:lam_nohard}}
\end{figure}

We have no strain hardening and therefore have only six internal variables $( \varepsilon_{11}^{p,\pm}, \varepsilon_{12}^{p,\pm}, \varepsilon_{22}^{p,\pm} )$ in the formulation.  We consider various combinations of material parameters that represent both small and large contrast, and the values are given in Table \ref{tab:lamparam}.  We generate data by picking average strain histories (applying the approach described in Section \ref{sec:arch} to each component of strain with a maximum strain of $33\%$) and computing the resulting stress histories.  We generate 800 trajectories, using 640 for training and 160 for testing.

Figure \ref{fig:lam_nohard} describes the results of training an RNO (default architecture changed to $4$ intermediate layers of $200$ nodes per layer) for the various cases.  Figure \ref{fig:lam_nohard}(a) shows the test error as a function of the number of internal variables for each of the choices of material parameters.  Here, we fix the number of internal variables, train the homogeneous material (MH), and then use transfer learning -- defined as using the trained RNO for MH as a starting point for training other models -- for the four cases (M1-M4) corresponding to different material parameters.  In each case, we see a sharp decline in error at two internal variables, and further (often more modest) drop in error when we increase from two to three internal variables.  Increasing the number of internal variables beyond three has little effect.  Further, the relative error, same as equation (\ref{eq:error}) in Section \ref{sec:1d}, is small, approximately 1-1.5\% relative error in each case.  Figure \ref{fig:lam_nohard}(b) compares the ``truth'' with the predictions of the trained RNO with three internal variables for the high contrast case M3.  Again, we see that the errors are very small.  The results are similar in other cases.

Figure \ref{fig:lam_nohard}(c-f) studies the ability to train an RNO with and without transfer learning.  In each of the cases M1-M4, we train the RNO with a fixed number of internal variables, and then transfer learning from the other cases.  In each case, direct training suggests the need for four internal variables.  However transfer-learning from the homogeneous case (MH) always leads to three internal variables.  This shows that the training can get stuck in local minima, a phenomenon 
which is not uncommon in neural networks in general, and may have more marked effect when
data volume is low 
\cite{kawaguchi2016deep,du2019gradient,li2018visualizing,safran2018spurious,rotskoff2022trainability}.  
Starting from the homogeneous case (MH) puts the RNO in a good basin and leads, in the experiments we have conducted, to a better local minimizer.

The approximation of the elasto-viscoplastic behavior by an RNO with three internal variables as demonstrated in Figure \ref{fig:lam_nohard}(a) (or four as in Figure \ref{fig:lam_nohard}(c-f)) is surprising in a system with six internal variables.  The possibility that the training process gets stuck in a local minimum would lead to a larger, not a smaller, number of internal variables.  Instead, the good approximation suggests that the equations of the system force the six state variables in the model ($\{ \varepsilon_{ij}^{p, \pm} (t) \}$ which we refer to as model variables in this section to distinguish them from the learnt internal variables of the RNO) to close proximity of a three dimensional manifold.  Similarly, notice that the test error decreases with the contrast MH-M3.  This is also surprising as one would expect larger fluctuations with higher contrast.  

We examine the range of values taken by the model variables using an auto-encoder to gain insight into these observations.  Briefly, an auto-encoder is a deep neural network approach to dimension reduction.  It is a deep neural network approximation of the identity map on some data in $\mathbb R^n$; it has a symmetric structure and an intermediate layer with a small latent dimension $m < n.$ 
In other words we may think of the auto-encoder as a map $A:{\mathbb R}^n \to {\ \mathbb R}^n$ such that $A = \psi_\text{AE} \circ \phi_\text{AE} \approx id$ on some set of data where $\phi_\text{AE}: {\mathbb R}^n \to {\ \mathbb R}^m, \psi_\text{AE}: {\mathbb R}^m \to {\ \mathbb R}^n$ are neural networks.  For any $x\in {\mathbb R}^n$, we call $\phi_\text{AE}(x) \in {\mathbb R}^m$ the reduced variable.  If we are able to successfully train an auto-encoder on the data, it means that for any data point, $x \approx \psi_\text{AE} ( \phi(_\text{AE}(x) )$; the reduced variable $\phi(x)$ ``encodes'' all the information in the data  \cite{lecun2015deep}.  In other words, the $n$-dimensional data, actually lives on a $m$-dimensional manifold.

\begin{figure}[t!]
    \centering
    \includegraphics[width=6in]{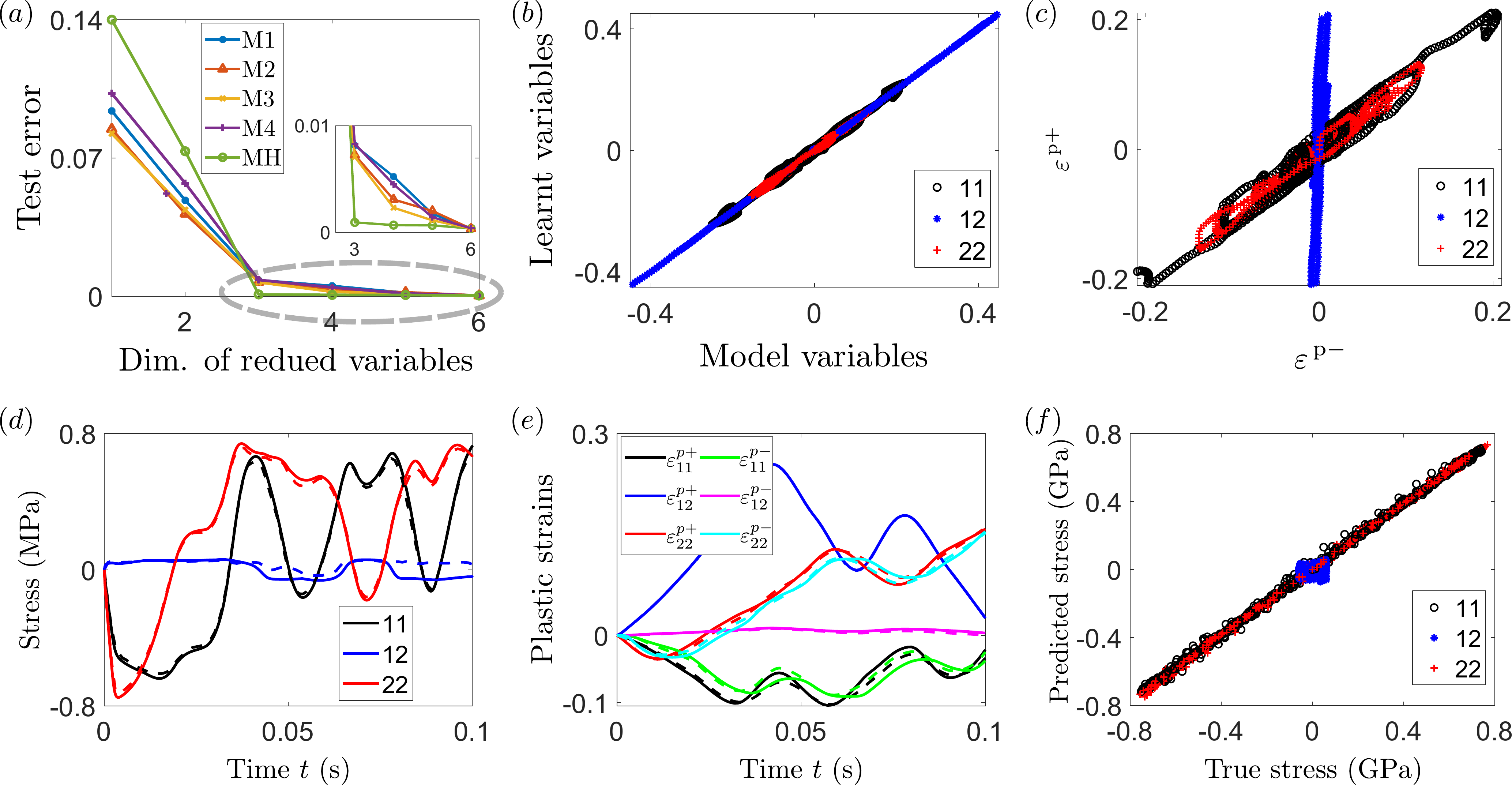}\\    
    \caption{Reducing the number of the model variables using an auto-encoder.  (a) The error of approximating the model variables (state variables in the original laminate model) with an auto encoder for various reduced dimensions.  (b)  Comparison between the computed values of the model variables and the recovered values from the trained auto-encoder for eight arbitrarily chosen trajectories in the test data for the case M3. (c) The computed values of the model variables in the the two layers for eight arbitrarily chosen trajectories in the test data for the case M3.  
    (d) Comparison of the computed stress response (solid) with the learnt stress response (dashed) for a typical test trajectory for the case M3.  (e) Comparison of the computed plastic strains model variables (solid) with the reconstructed plastic strain  (dashed) for a  typical test trajectory for the case M3.  (f) Comparison of the computed stress with the learnt stress for eight test trajectories for the case M3.
    }
    \label{fig:lamae_nohard}
\end{figure}

We take each of the trajectories and take the values of the model variables at each computed instant $t$,  $\{ \varepsilon^{p, \pm} (t) \}$, as a six-dimensional data point. Since we generate 800 trajectories each with 100 time-steps, we have a rich data set of $8\times 10^4$ data points.  We use $6.4\times 10^4$ (corresponding to 640 trajectories) of these for training an auto-encoder, consisting of one layer with 50 nodes in both the encoding and decoding portions, and $1.6\times 10^4$ for testing.  We train auto-encoders with varying reduced dimension.  The results are shown in Figure \ref{fig:lamae_nohard}.  We see that in Figure \ref{fig:lamae_nohard}(a) that for all cases, the results are well approximated with a reduced dimension of three using a relative reconstruction error,
\begin{equation}
\text{error} = \frac{1}{S} \ \sum_{s=1}^S \frac{ \lvert \varepsilon_s^{p\pm,\, \text{model}} - \varepsilon_s^{p,\, \text{learned}} \rvert}{  \lvert \varepsilon_s^{p\pm, \, \text{model}} \rvert }
\end{equation}
where $s$ indexes the $8\times 10^4$ data points and $|\cdot|$ denotes an $l^2$ norm of the six-dimensional data points $\{ \varepsilon^{p, \pm} (t) \}$.  Figure \ref{fig:lamae_nohard}(b) compares the actual values of the model variables and those that are reconstructed from the reduced variables.  We observe good reconstruction indicating that the auto-encoder is able to correctly encode the model variables.  All of this leads us to the conclusion that the equations of the system force the six model variables to (the close proximity) of a three dimensional manifold in all cases.  This agrees with what we anticipated from the RNO.   

We explore this further in Figure \ref{fig:lamae_nohard}(c) that compares the values of the model variables (plastic strain) in the $+$ layer with that in the $-$ layer.  It shows that the $11$ and $22$ components of the plastic strain are almost equal in both layers while the $12$ component in the (harder) $-$ remains relatively small.  This observation provides us with a heuristic insight as to why we only need three internal variables in the RNO approximation.  When the imposed strain is small, the response is largely elastic and therefore one does not need internal variables.  When the imposed strain is large (so that the plastic strain is much larger than the elastic strain), the strain compatibility (\ref{eq:jump})$_1$ forces the $22$ and $33$ components of strain to be approximately equal, $\varepsilon_{22}^{p, +} \approx \varepsilon_{22}^{p,-}, \varepsilon_{33}^{p, +} \approx \varepsilon_{33}^{p,-}$.  Since the plastic strain is trace-free,  $\varepsilon_{33}^{p,\pm} = - \varepsilon_{11}^{p,\pm} - \varepsilon_{22}^{p,\pm} $, and we conclude that the $11$ component of the plastic strains are also equal, $\varepsilon_{11}^{p, +} \approx \varepsilon_{11}^{p,-}$.   Furthermore, the stress continuity condition (\ref{eq:jump})$_2$ means that the $11, 12$ components are the same and that the stress can only differ in the $22$ component.  This limits the fluctuation in the deviatoric stress, and we expect plastic yield in the softer layer (+ layer) to dominate over the other.  Since only the $12$ components of plastic strain can differ from one another, we expect the value in the softer $+$ material to be significantly larger than that in the harder $-$ material, $\varepsilon_{12}^{p,+} >> \varepsilon_{12}^{p,-}$.   In short, at large strains, we only need information from the softer layer, and thus only three internal variables.  Further, this approximation improves when we have higher contrast.


The fact that the reduced variables encode the model variables suggests that they can be used as internal variables in an RNO.  So we again train an RNO of the form (\ref{eq:RNO}) with three internal variables ($k=3$) set to be the reduced variables learnt by the auto-encoder.  To elaborate, let $\phi_\text{AE}: \varepsilon^{p,\pm} \to \zeta$ denote the trained auto-encoder from the model variables $\varepsilon^{p,\pm} \in {\mathbb R}^6$ to the reduced variables $\zeta \in {\mathbb R}^3$.  So, for any trajectory, at any instant, we use $\{ \bar{\varepsilon}(t), \phi_\text{AE}(\varepsilon^{p, \pm}) (t) \}$ as input and $\{\bar{\sigma}(t), \dot {\overline{\phi_\text{AE}(\varepsilon^{p,\pm})}}(t)\}$ as output to train a neural network approximation for the map 
\begin{equation} \label{eq:aerno}
\Phi: \begin{cases}
\sigma = f (\bar{\varepsilon}, \zeta),\\
\dot{\zeta} = g (\bar{\varepsilon}, \zeta).
\end{cases}
\end{equation}
We train this map on a training data set consisting of 100 time-steps across 800 trajectories, with an architecture of $6$ layers and $400$ nodes, and test it against a test data set consisting of 100 time-steps across 200 trajectories.  The overall error is 2.39\%.  Figure \ref{fig:lamae_nohard}(d) compares the actual computed stress response (solid line) with the learnt response (dashed line) for a typical test trajectory.  Figure \ref{fig:lamae_nohard}(e) compares model variables (solid line) with those reconstructed from the learnt map (\ref{eq:aerno}) and the auto-encoder for a typical test trajectory.  Specifically, for a given strain trajectory, we predict the stress $\sigma$ and $\zeta$ using the learnt map (\ref{eq:aerno}), and then reconstruct the plastic strains from the learnt auto-encoder $\psi_\text{AE} (\zeta)$. Figure \ref{fig:lamae_nohard}(f) repeats the stress comparison for eight trajectories chosen arbitrarily from the test data.

\begin{figure}[t]
    \centering
    \includegraphics[width=4in]{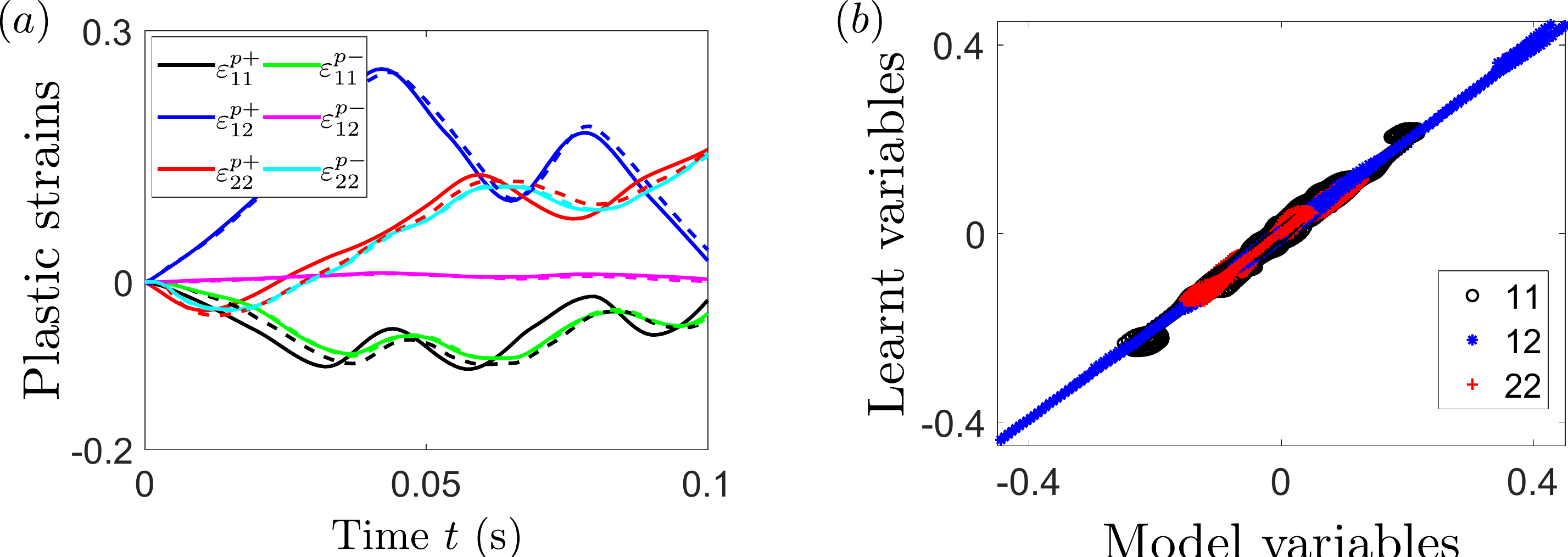}
    \caption{Reconstructing the micromechanical model variables from the macroscopic internal variables for the case M3.  Comparison between the computed values of the model variables and the recovered values from the RNO internal variables for (a) an arbitrarily chosen trajectory and (b) eight arbitrarily chosen trajectories.}
    \label{fig:lam_nohard2}
\end{figure}

Recall from Section \ref{sec:arch} that the internal variables have no inherent meaning, and that reparametrization can lead to an equivalent set of internal variables.  In the passage above, we identified one set of internal variables by training the RNO purely from the macroscopic strain trajectory to stress trajectory map, and an alternate set by reducing the dimension of the microscopic or model variables using an auto-encoder.   The fact that both sets can approximate the macroscopic response suggests that they are equivalent sets of internal variables.  We conclude this section by studying the converse.  Figure \ref{fig:lam_nohard2} shows that the internal variables learnt by training the RNO purely from the macroscopic strain trajectory to stress trajectory response does in fact reconstruct the micromechanical model variables.   We train a neural network from the internal variables to the model variables, and the figure shows that it provides a good representation.  We should note, however, that this ability of the internal variables to learn the model variables may be very particular to the case of the laminate -- in general the map from the micromechanical variables in high dimension to the macroscopic internal variables in low dimension may not be invertible. 

\subsection{Exponential strain hardening} \label{sec:lam_exphard}

\begin{figure}[t]
    \centering
    \includegraphics[width=6in]{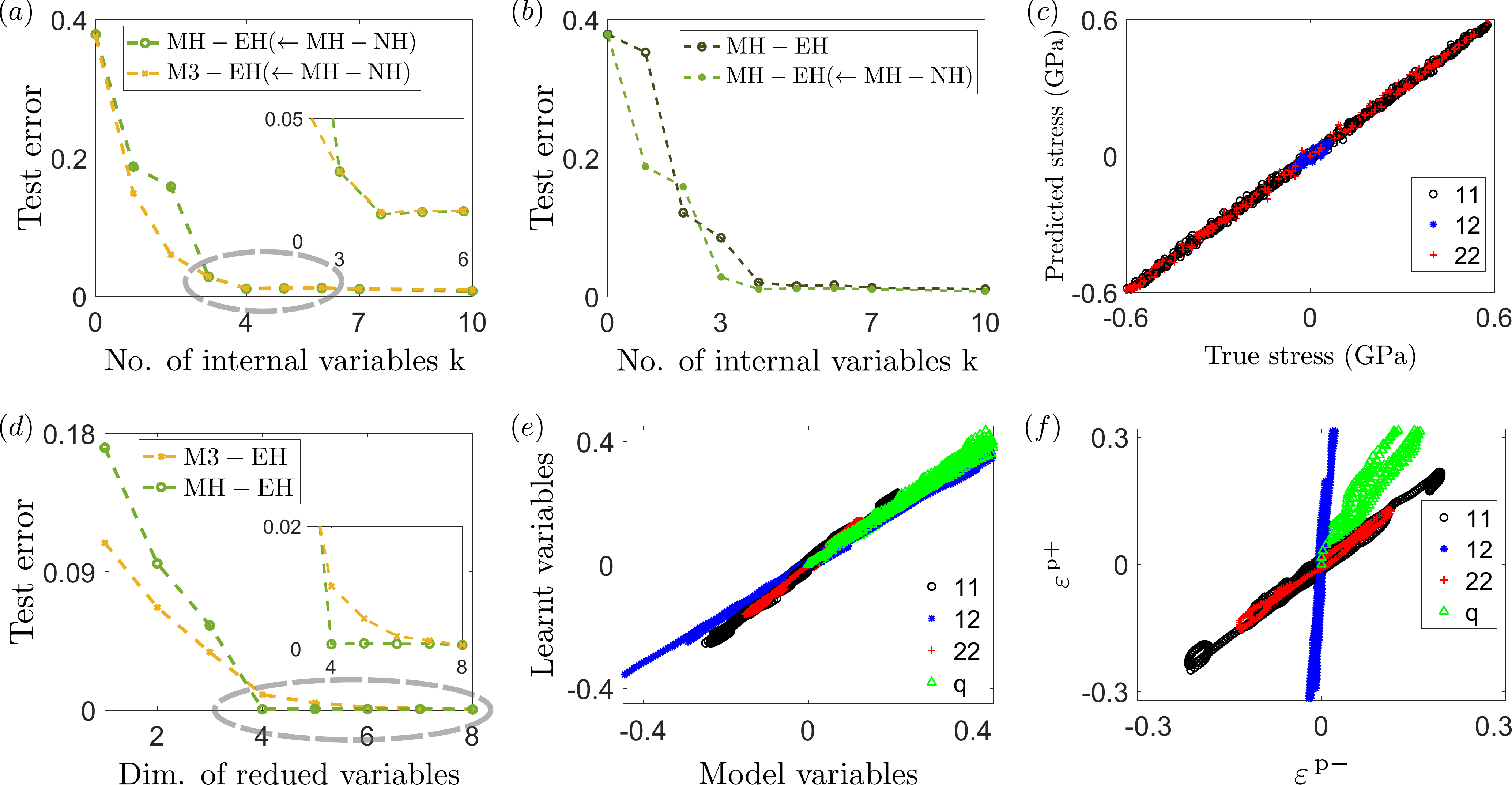}
    \caption{ Approximating an elasto-viscoplastic laminate with exponential strain hardening. (a-c) Learning the strain history to stress history map with a recurrent neural operator: (a) Training RNOs with various numbers of { internal} variables. In each case, we train the homogeneous material with no hardening (MH-NH) and then use transfer learning. (b) Training RNOs with various numbers of { internal} variables both directly and with transfer learning. (c) Comparison of the ground truth and RNO output for eight test strain trajectory for the case M3-EH. (d-f) Reducing the number of the internal variables using an auto-encoder: (d) The error of approximating the model variables (state variables in the original laminate model) with an auto encoder for various reduced dimensions. (e)  Comparison between the computed values of the model variables and the recovered values from the trained auto-encoder for eight arbitrarily chosen trajectories in the test data for the case M3-EH. (f) The computed values of the model variables in the the two layers for eight arbitrarily chosen trajectories in the test data for the case M3-EH.
    \label{fig:lam_hard}}
\end{figure}

We repeat the analysis with exponential strain hardening.  Figure \ref{fig:lam_hard}(a-c) shows the results of training a RNO for the homogeneous case (MH), and the case (M3) with an exponential hardening coefficient of $h=1$ (MH-EH and M3-EH).  We see in Figure \ref{fig:lam_hard}(a) that we need four internal variables (in contrast to the case of no hardening where we needed only three). Furthermore, we observe that the final error is larger, though we do obtain good approximation. Figure \ref{fig:lam_hard}(b) shows that transfer learning (as defined in the preceding subsection) starting from the homogeneous material with no hardening leads to a better approximation even for the homogeneous case. Figure \ref{fig:lam_hard}(c) compares the ground truth and RNO output for eight test trajectories for the case (M3-EH).  Figure \ref{fig:lam_hard}(d-f) explores the use of an auto-encoder to learn the reduced variables from the eight model variables.   We see in Figure \ref{fig:lam_hard}(d) we need four reduced variables consistent with the RNO, and in Figure \ref{fig:lam_hard}(e) that these are able to correctly recover the model variables. Figure \ref{fig:lam_hard}(f) compares the model variables in one layer with that in the other.  As before, we see that the $11$ and $22$ components of the plastic strain are almost equal, and the $12$ component  in the $+$ layer dominates over that in the $-$ layer (same argument holds).  The hardening variable $q$  in the $+$ layer also dominates, but not as much; still, the values in the two layers are proportional to each other, meaning that we need only one value to describe the overall behavior.
\vspace{0.1in}

In summary, we find that the RNO trained only on macroscopic stress-strain data provides an accurate approximation of the actual response of a laminated composite.  Further, the training process shows that we require half the number of internal variables than one would expect, and an analysis of the micromechanical fields shows that the governing equations force the micromechanical variables to the proximity of a low dimensional manifold.

\section{Elasto-viscoplastic polycrystals} \label{sec:learn}

We now consider a polycrystalline medium that is an ensemble of grains, each made of the same anisotropic material but with varying orientations.  We assume that each grain is elasto-viscoplastic with a number of slip and twin systems. We adopt the two scale framework introduced in Section \ref{sec:arch}, generate data using an unit cell problem and then use this data to train the RNO approximation in both two and three dimensions.

\subsection{Two dimensions with data generated by full-field solution} \label{sec:2}

The data in two dimensions is generated from a polycrystalline unit cell with 32 randomly oriented grains generated using periodic Voronoi tessellation \cite{fritzen2009periodic} according to the microscopic model described in \ref{sec:app}.  The corresponding full-field unit cell problem with two slip systems is solved using the accelerated computational micromechanics framework \cite{zhou_accelerated_2021}. We consider a system with two slip systems with parameters shown in Table \ref{tab:2D} unless otherwise specified.

\begin{table}
{
\centering
\begin{tabular}{cccccccc}
\hline
\multicolumn{8}{c}{Elastic: all cases}\\
\multicolumn{8}{c}{$\mu = 19$ GPa, $\kappa = 48$ GPa}\\
\hline
\multicolumn{8}{c}{Slip system 1}\\
 Case& $b_1$& $n_1$& $\tau^\text{p}_{0,1}$ (MPa)& $\dot{\gamma}^\text{p}_{0,1}$ (1/s)& $m_1$& $\sigma_1^\infty$ (MPa) & $h_1$ (MPa)\\
PH,P1,P2 & (1, 0) & (0, 1) & $100$ & $1$ & $0.05$ & $2$ & $7100$\\
P3& (1, 0) & (0, 1) & $100$ & $1$ & $0.05$ & $50$ & $100$\\
P4& (1, 0) & (0, 1) & $100$ & $1$ & $0.05$ & $100$ & $100$\\
P5& (1, 0) & (0, 1) & $100$ & $1$ & $0.05$ & $100$ & $100$\\
\hline
\multicolumn{8}{c}{Slip system 2}\\
Case & $b_2$& $n_2$& $\tau^\text{p}_{0,2}$ (MPa)& $\dot{\gamma}^\text{p}_{0,2}$ (1/s)& $m_2$& $\sigma_2^\infty$ (MPa) & $h_2$ (MPa)\\
PH& (1, 1) & (-1, 1) & $100$ & $1$ & $0.05$ & $2$ & $7100$\\
P1& (1, 1) & (-1, 1) & $100$ & $1$ & $0.25$ & $2$ & $7100$\\
P2& (1, 1) & (-1, 1) & $100$ & $1$ & $0.5$ & $2$ & $7100$\\
P3& (1, 1) & (-1, 1) & $100$ & $1$ & $0.05$ & $50$ & $100$\\
P4& (1, 1) & (-1, 1) & $100$ & $1$ & $0.05$ & $100$ & $100$\\
P5& (1, 1) & (-1, 1) & $500$ & $1$ & $0.25$ & $500$ & $100$\\
\hline
\end{tabular}
\caption{Material parameters used for polycrystals in two dimensions. \label{tab:2D}}
}
\end{table}

Recall that we take the elastic behavior to be the compressible neo-Hookean (Appendix \ref{eq:elas}) at the microscale, and it is easy to verify that the Cauchy stress may be decomposed into hydrostatic and deviatoric components
\begin{equation}
\sigma^{\mu} (F^{\mu},F^{p,\mu}) = - p (\det F^{\mu}) I + \sigma^{\text{dev},\mu}(F^\mu, F^{p,\mu})
\end{equation}
where we use the superscript $\mu$ to emphasize that this holds at the microscopic scale, and the fact that $\det F^{p,\mu}=1$.  It follows that at the macroscopic scale,
\begin{equation}
\sigma = - \langle - p (\det F^{\mu}) \rangle I + \langle \sigma^{\text{dev},\mu}(F^\mu, F^{p,\mu}) \rangle = - p I + \sigma^\text{dev}.
\end{equation} 
Now, since $\det F^{p,\mu}=1$, the pressure at any point in the unit cell is independent of the plastic strain at that point. So, it is independent of the history of that point.  However, the overall stress distribution and therefore the overall hydrostatic pressure can depend on the distribution of the plastic strain in the unit cell.  Therefore, it is possible the overall hydrostatic pressure can develop history dependence from the evolution of the plastic strain distribution. Still, we expect that the deviatoric stress would be much more sensitive to the history. Therefore, we fit the hydrostatic pressure and deviatoric stress separately: in other words, we train an RNO to learn the hydrostatic pressure and another independent RNO to learn the deviatoric stress\footnote{Or, one can think of them as one RNO with two disjoint parts.}.

\begin{figure}[t!]
    \centering
    \includegraphics[width=6in]{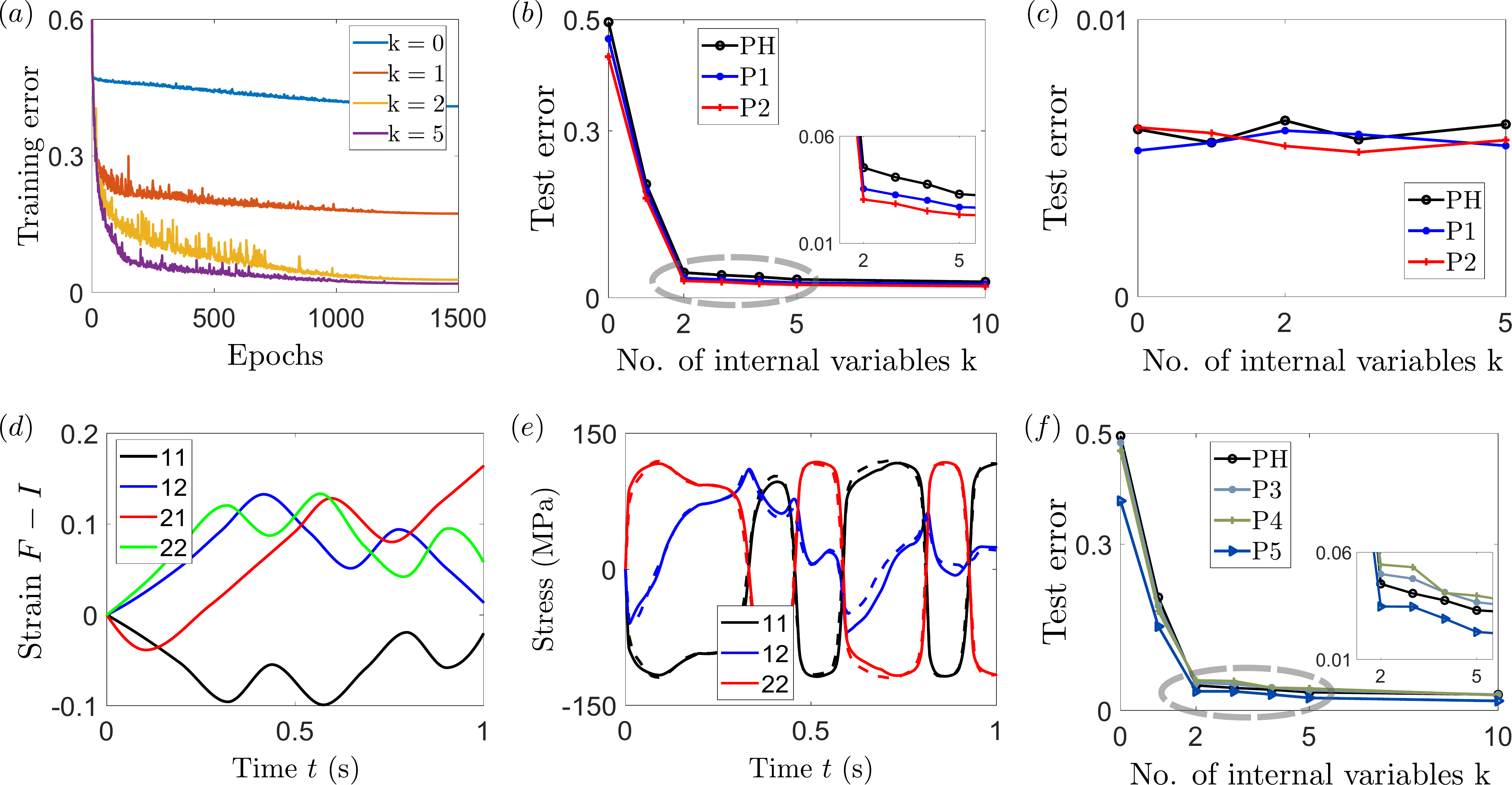}
    \caption{Approximating two dimensional polycrystalline elasto-viscoplasticity with a recurrent neural operator. (a) Training RNOs with various numbers of { internal} variables.  (b) Test error of the trained RNOs of the deviatoric stress with various number of { internal} variables and rate hardening cases. (c) Test error of the trained RNOs of the hydrostatic pressure with various number of { internal} variables and rate hardening cases.  (d) A typical average deformation gradient history input test data.  (e) Comparison of the resulting deviatoric stress history for the test data (solid) and the prediction of the RNO with two { internal} variables (dashed). (f) Test error of the trained RNOs of the deviatoric stress with various number of { internal} variables and strain hardening cases.
    \label{fig:2DFFT}}
\end{figure}

We generate data associated with 500 trajectories for various combinations of material parameters.  We use 400 trajectories for training and 100 trajectories for testing.  Again, for each case we train the model with various numbers of internal variables using $4$ nodes fo $200$ nodes each.  The results are shown in Figure \ref{fig:2DFFT}.  Figure \ref{fig:2DFFT}(a) shows the decrease of the training error for the deviatoric stress with the number of training epochs for the case with parameters in Table \ref{tab:2D}.  We see that the error decreases rapidly and we obtain convergence with about 1500 epochs. The results are similar in other cases.

Figure \ref{fig:2DFFT}(b) shows the test error in learning the deviatoric stress with for the PH and P1-P3 material cases shown in Table \ref{tab:2D}.  In each case, we see that the RNO with no internal variables has a large error, but the error reduces with one and two internal variables. Increasing the number of internal variables beyond two does not reduce the error.  Moreover, the relative error is small.  Thus, we conclude that the data is well-represented with two internal variables.  Figure \ref{fig:2DFFT}(c) shows the corresponding results for the hydrostatic pressure: we see no dependence of the error on the the number of internal variables, and the error is small.   Therefore, we conclude that the hydrostatic pressure has no history dependence.  Figure \ref{fig:2DFFT}(d) shows a typical deformation gradient trajectory, and Figure \ref{fig:2DFFT}(e) compares the `truth' vs. the learned corresponding stress trajectory for the PH material case. Note that this is the only the deviatoric stress. Again, we see that the learned response provides a very good approximation for the data.  Finally, we study the effect of strain hardening in Figure \ref{fig:2DFFT}(f): we again observe that we obtain good approximation with two internal variables, and increasing the number of internal variables beyond two does not provide any better approximation.

In summary, we find that the RNO is able to provide a highly accurate surrogate for 2D crystal plasticity.  Further, we only require two state variables for various choices of parameters.   In two dimensions, deviatoric stresses as well as isochoric stretches associated with plasticity lie on a two-dimensional manifold.  The fact that we are able to represent the stretch vs. stress trajectory with two hidden variables is therefore significant.  At the same time, we do not know how general this is, and discuss this further in Section \ref{sec:disc}.

}

\subsection{Three dimensions with data generated by Taylor averaging} \label{sec:3D}

The data in three dimensions is generated from a polycrystalline unit cell with 128 randomly oriented grains using Taylor averaging where the solution to the equilibrium equation is replaced by the assumption that the deformation gradient is uniform \cite{ketal_book_00}.
 Each grain has three basal, three prismatic and six pyramidal slip systems in addition to six tensile twin systems (treated as pseudoslip), and the material parameters chosen to represent magnesium  \cite{Chang2015AMagnesium}.   We refer the reader to \cite{liu2022learning} for details\footnote{We take this opportunity to correct an incorrect statement there.  The implementation by  \cite{Chang2015AMagnesium} does not compute yield, but updates the slip and psuedo slip from zero stress.}.

\begin{figure}[t!]
    \centering
    \includegraphics[width=5in]{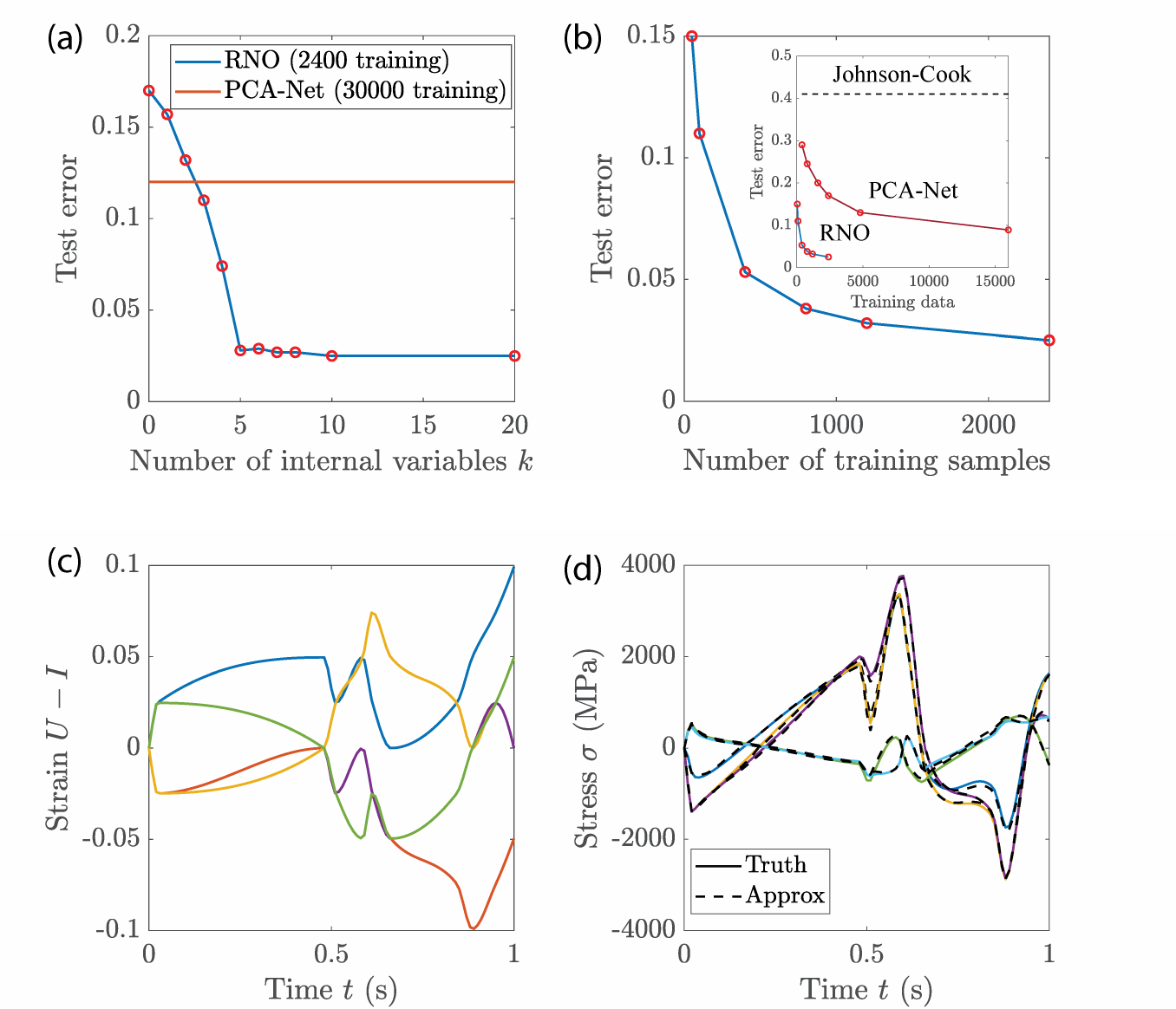}
    \caption{Approximating three dimensional polycrystalline elasto-viscoplasticity with a recurrent neural operator. (a) Test error of the trained RNOs of a various number of { internal} variables; also shown in the dashed line is the test error of a PCA-Net.  (b) Test error as a size of the training data.   The inset shows the corresponding test error for the PCA-Net and Johnson-Cook model fitted in a classical manner.       (c) A typical average deformation gradient history input test data.  (d) Comparison of the resulting stress history for the test data (``Truth'') and the prediction of the RNO with five { internal} variables (``Approx'').} 
    \label{fig:3D_test}
\end{figure}

We use up to 2400 realizations to train RNOs with 0 through 10 { internal} variables. The architecture, optimizer and hyper parameters  are the same as in one dimension described in Section \ref{sec:1d}, except the number of internal layers for both of $f$ and $g$ are increased to 5, and the width of each internal layer is increased to 200. 

The results are shown in Figure \ref{fig:3D_test}.  Figure \ref{fig:3D_test}(a) shows the test error of the RNOs with varying number of { internal} variables tested against 25000 realizations.  We see that the RNO with no { internal} variables has large error, but the error reduces with increasing number of { internal} variables until it saturates at five { internal} variables.   Increasing the number of { internal} variables beyond five does not reduce the error.  Thus, we conclude that the data is well-represented with five { internal} variables.  Figure \ref{fig:3D_test}(b) shows the test error as a function of the size of the training dataset which enables us to conclude that 2400 trajectories provides excellent representation.  A typical input deformation gradient history and output stress history (both the original data (``Truth'')  and the prediction of the RNO with five { internal} variables (``Approx'')) are shown in Figures \ref{fig:3D_test}(c) and (d) respectively.

In three dimensions, deviatoric stresses and isochoric stretches associated with plasticity lie on a five-dimensional manifold.  The fact that we are able to represent the stretch vs. stress trajectory with five { internal} variables is therefore significant.   { At the same time, we do not know how general this is, and discuss this further in Section \ref{sec:disc}.}

In Figure \ref{fig:3D_test}(a), we also show the test error associated with a PCA-Net trained with 30000 trajectories \cite{liu2022learning}.  We observe that the RNO with five internal variables performs significantly better than the PCA-Net trained with a significantly larger data set.  This is also reflected in the inset of Figure \ref{fig:3D_test}(b) that compares test error vs size of the training data for both the current RNO and the previous PCA-Net.  

\begin{figure}
    \centering
    \includegraphics[width=6.5in]{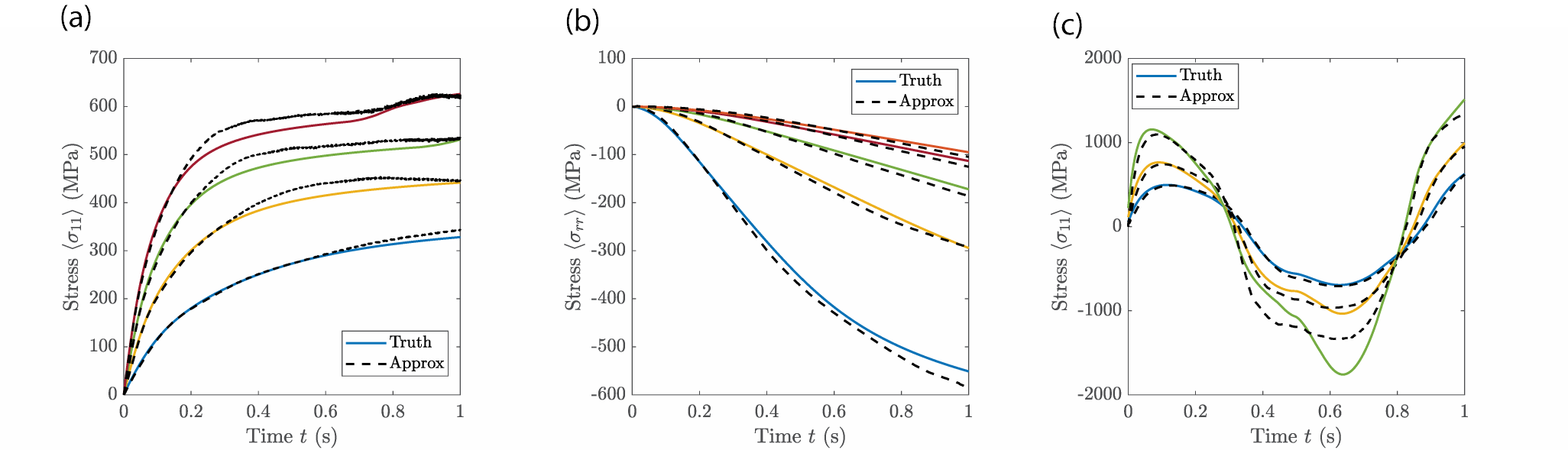}
    \caption{ Testing the trained RNO against classical loading paths. (a) Uniaxial tension. (b) Cavity expansion. (c) cyclic loading.  }
    \label{fig:3D_classical}
\end{figure}

We further test the efficacy of our trained RNO with five internal variables with three loading paths that are traditionally used to train material models: uniaxial tension, cavity expansion and uniaxial cyclic loading.  The results, shown in Figure \ref{fig:3D_classical}, show good ability to reproduce these tests despite the fact that these trajectories are not included in the training data.  Further, the resulting error of 8\% is better than the error of the best fit Johnson-Cook material model \cite{johnsoncook} that is widely used and fit against this very data.   Indeed, testing the best fit Johnson-Cook material model that is fit against these classical tests produces an average error of 40\% when tested against the random trajectories used in this work (see inset of Figure \ref{fig:3D_test}(b).

In summary, we find that the RNO is able to provide a highly accurate surrogate for 3D crystal plasticity.  Further, the RNO requires significantly less data to train compared to our previous work on PCA-net, and results in a smaller error.  Finally, the RNO provides a resolution independent representation.

\section{Resolution independence and architecture}  \label{sec:res}

\begin{figure}
    \centering
    \includegraphics[width=6.5in]{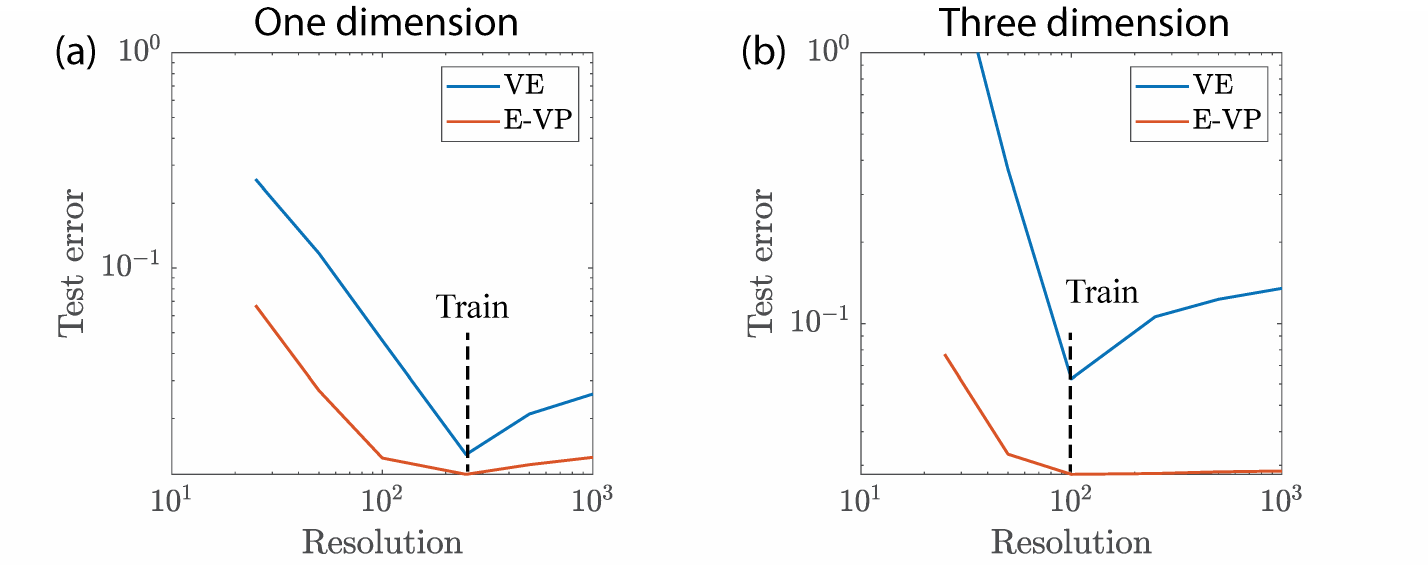}
    \caption{Resolution independence and architecture.  Comparison of the test error of an RNO trained with training data at a resolution of 200 time-steps again test data various resolution, for both the elasto-viscoplastic RNO (\ref{eq:RNO}) labelled ``E-VP'' and the alternate viscoelastic RNO (\ref{eq:RNO2}) labelled ``VE'' .  (a) One-dimensional elasto-viscoplastic composite of Section \ref{sec:1d}, (b) Three-dimensional elasto-viscoplastic polycrystal of Section \ref{sec:3D}.}
    \label{fig:3D_res}
\end{figure}

As noted in the introduction, it is desirable that any neural operator approximation be independent of the resolution so that it represents the actual map rather than the numerical data that is used to train it.  We test this in Figure \ref{fig:3D_res}.  In the  examples studied in Sections \ref{sec:1d} and \ref{sec:3D}, we solve the elasto-viscoplastic unit cell problem using 1000 time steps.  We down-sample it to 200 time steps and use this as our training data.    We then test the RNO trained at 200 time steps against test data down sampled to various resolutions.  We see in Figure \ref{fig:3D_res} that the error elasto-viscoplastic RNO (``E-VP'') is largely independent of the resolution in all the three examples that we have studied.  Thus, the RNO (\ref{eq:RNO}) provides a resolution-independent representation of elasto-viscoplasticity. 

We then examine the alternate viscoelastic  RNO (\ref{eq:RNO2}), where the stress depends on both stretch and stretch rate, on the same elasto-viscoplastic data.  We again train it with data at a resolution of 200 time steps in each example and test it against test data at various resolution.  We observe from Figure \ref{fig:3D_res} that while the test error is small at the training resolution, it is large at other resolutions.  Thus, the RNO (\ref{eq:RNO2}) does {\it not} provides a resolution-independent representation of elasto-viscoplasticity. 

This is interesting for three reasons.  First,  the elasto-viscoplastic RNO (\ref{eq:RNO}) may be regarded as a special case of the alternate viscoeleastic RNO (\ref{eq:RNO2}).  Therefore, the viscoelastic RNO (\ref{eq:RNO2}) has all the ability of the elasto-viscoplastic RNO (\ref{eq:RNO}).  However, the extra freedom leads to over-training at the training resolution, and is unable to provide the right representation for processes at a different resolution.  Second, we have studied viscoelasticity in the companion paper \cite{viscoelasticity}, and there the situation is reversed: the viscoelastic RNO (\ref{eq:RNO2}) provides resolution-independent representation while the elasto-viscoplastic RNO (\ref{eq:RNO}) does not (it has low error at the training resolution and not elsewhere).   In viscoelasticity, one needs the extra freedom of the viscoelastic RNO (\ref{eq:RNO2}) to have resolution independence.  Finally, from a physical point of view, note that the crystal at the individual grain does not have any viscoelasticity (explicit rate-dependence in the stress-stretch relationship).  The good resolution-independence of the the elastic-viscoplastic RNO (\ref{eq:RNO}), and the poor resolution-independence of the the viscoelastic RNO (\ref{eq:RNO2}) shows that the polycrystal does not acquire any viscoelasticity from the averaging process and the system remains elastic-viscoplastic.

Together all of this shows that there is a connection between the underlying physics, architecture, training and resolution independence.   While an architecture that does not necessarily represent the correct physics may be trained at a particular resolution, it fails when used at another.  Thus examining RNOs at various resolution is a useful tool to understand both the efficacy of the approximation and aspects of the underlying physics.

\section{Multiscale simulation}  \label{sec:multi}

The goal of developing a machine-learned surrogate constitutive model is to use it in macroscopic calculations, and we do so here for two classical impact problems in three dimensions.  We implement the trained RNO with five internal variables described in Section \ref{sec:3D} as a material model in (``VUMAT'') in the commercial finite element package ABAQUS \cite{manual2014abaqus} and perform various tests.  This implementation exploits the local nature of the RNO architecture.  This enables us to directly calculate the stress and internal variable update based on the deformation gradient update and current value of the internal variable following (\ref{eq:RNOd}).   We emphasize that the RNO is trained only once for all the calculations presented below; the cost of training is amortized across all future uses of the machine-learnt surrogate constitutive model.  We use the first example to show how one can use the RNO for a design parameter study and the second example to demonstrate resolution independence.

 \subsection{Projectile impact on a plate}

\begin{figure}[t]
    \centering
    \includegraphics[width=5in]{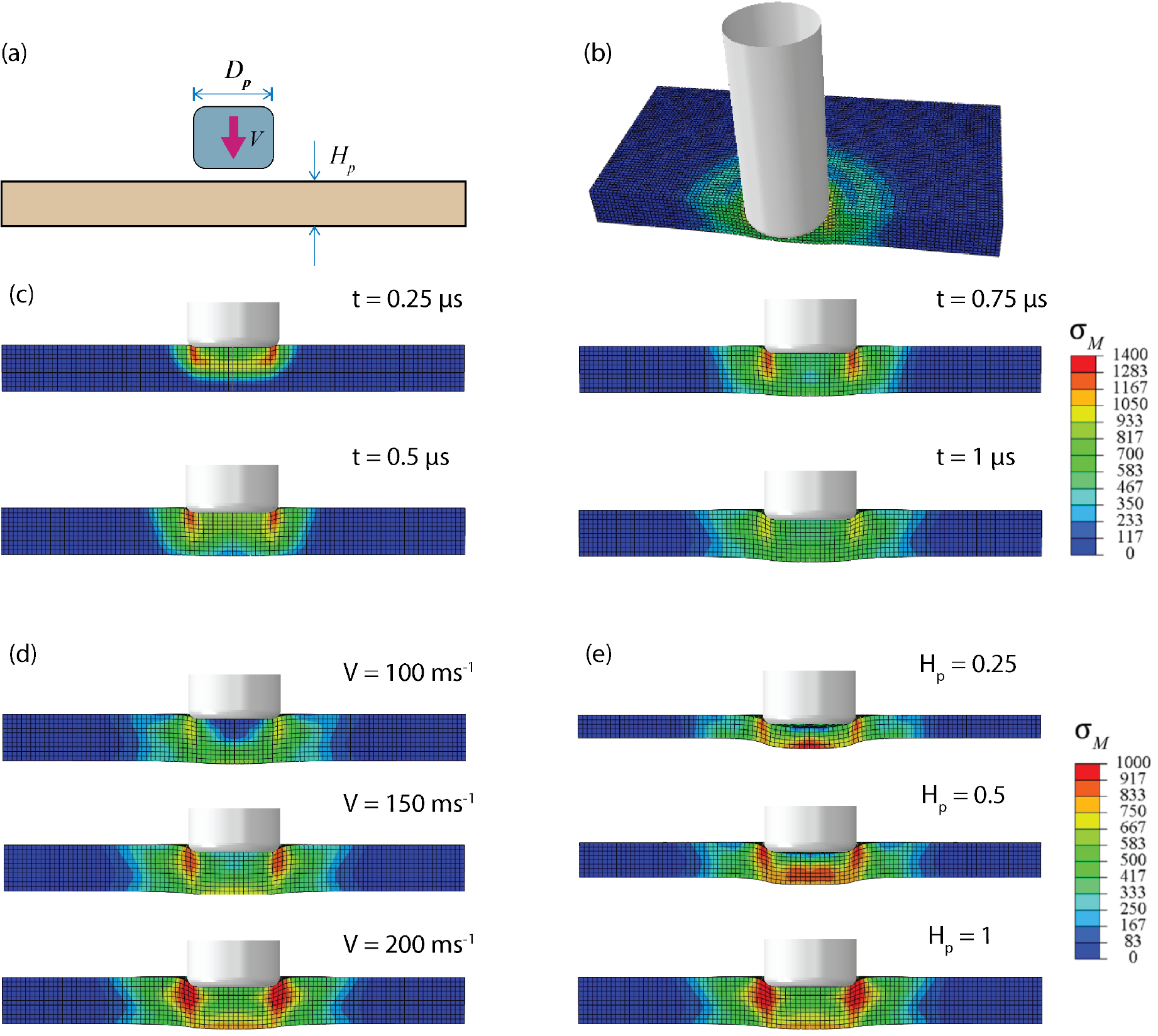}
    \caption{Projectile impact on a plate. (a) Schematic.  (b) Deformation in three dimensions.  (c) Snapshots an axial cross-section with the von Mises stress measure. (d)  Axial cross-section with von Mises stress measure at $t=1s$ for various impact velocities $V$.  (e)  Axial cross-section with von Mises stress measure at $t=1s$ for various plate thicknesses $\hat{H}_p=H_p/D_p$. }
    \label{fig:plate}
\end{figure}

The first test concerns the impact of a blunt rigid heavy projectile on a rigid plate as shown schematically in Figure \ref{fig:plate} (a).   
Figures \ref{fig:plate}(b) and (c) show the result of a typical test where a cylindrical projectile of radius $D_p$ = 2 mm, traveling at a velocity $V = $ 200 m/s, impacts a large magnesium plate of thickness $H_p$ = 1 mm which is simply supported far away from the point of impact.   The von Mises strain $\sigma_M = \sqrt{3}/2 | \sigma - (\text{tr }\sigma) / 3 \  I |$ is shown in Figure \ref{fig:plate}(c).  An elastic wave followed by a plastic wave propagates into the plate and is reflected from the free-face.  Subsequently, the plate becomes a wave guide with radially expanding elastic and plastic waves as the impactor penetrates into the plate.   Figures \ref{fig:plate} (d) and (e) show the parametric study for various impact velocities (d) and plate thicknesses (e).  These results agree qualitatively with those of our prior results \cite{liu2022learning} using a PCA-Net as VUMAT, though the actual stress distributions differ slightly due to differences in the ability to approximate the polycrystal behavior.

\subsection{Taylor impact test}

\begin{figure}[t]
    \centering
    \includegraphics[width=5in]{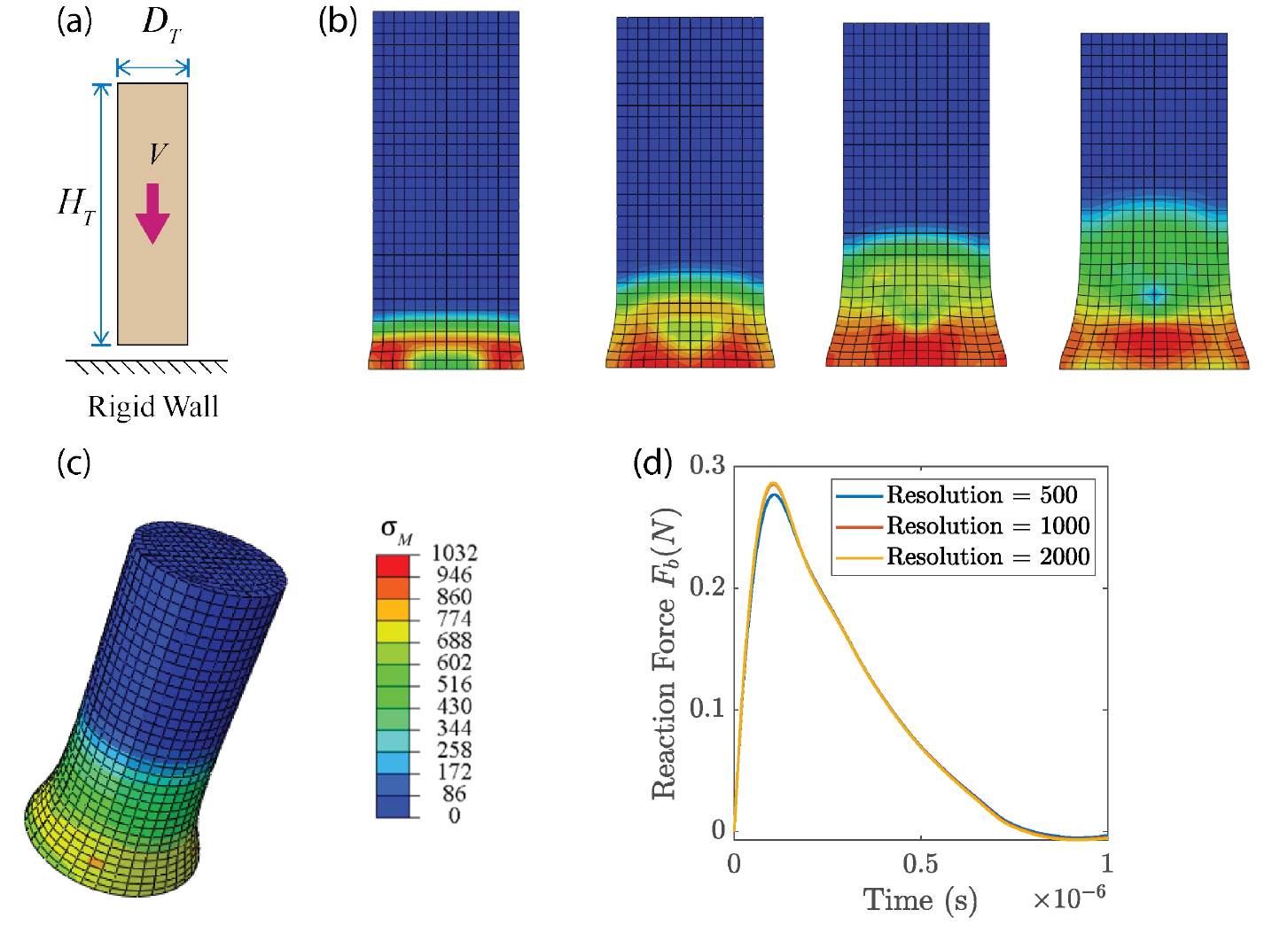}
    \caption{Taylor impact. (a) Schematic.  (b) Deformation in three dimensions.  (c) Snapshots an axial cross-section with the von Mises stress measure. (d) The reaction force vs. time for various time resolution.}
    \label{fig:impact}
\end{figure}

The second is the Taylor impact test shown in Figure \ref{fig:impact}.  A magnesium cylinder of height $H_T=5$mm and diameter $D_T=1$mm, traveling with an initial velocity $V=200$m/s, impacts a rigid friction-less wall at time $t=0$ as shown schematically in Figure \ref{fig:impact}(a).  Figure \ref{fig:impact}(b) shows a snapshot in perspective view while Figure \ref{fig:impact}(c) shows various snap-shots of the cross-section (the color denoting the von Mises stress).    Upon impact, an elastic wave first propagates into the impactor followed by an expanding region of plastic deformation.  We also have a release wave from the side leading to a complex radial distribution of the plastic deformation.  These results agree qualitatively with those of our prior results   \cite{liu2022learning} using a PCA-Net as VUMAT, though the actual stress distributions differ slightly due to differences in the ability to approximate the polycrystal behavior.

We use the example to demonstrate the (time) resolution-independence of the RNO, and convergence with time-step.  We repeat the calculation with various time resolution, and we find in Figure \ref{fig:impact}(d) that the results, and specifically the reaction force, remain essentially unchanged.  The peak reaction force is slightly lower for the coarsest resolution as expected.

\subsection{Computational cost} 

\begin{figure}
    \centering
    \includegraphics[width=4in]{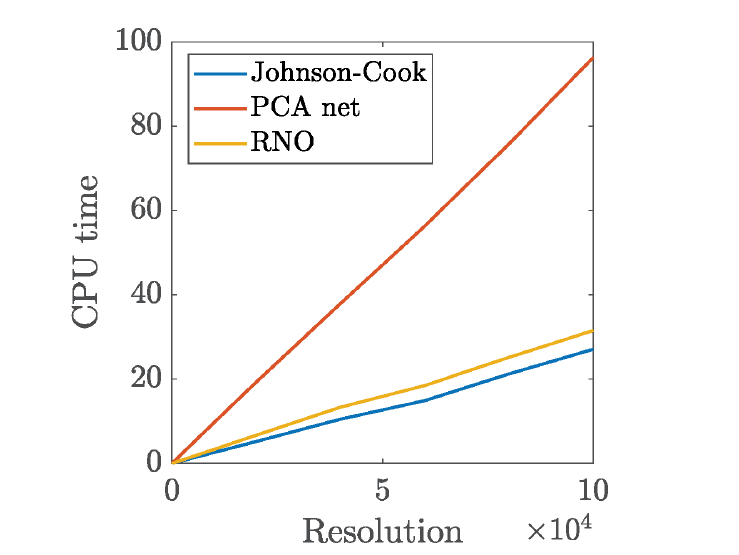}
    \caption{ Computational cost.: CPU time versus number of time steps for the classical Johnson-Cook constitutive law, the PCA-Net proposed in our previous work and the current RNO.}
    \label{fig:cost}
\end{figure}

We finally assess the computational cost of using the trained RNO in a finite element calculation.  We conduct a single element calculation in ABAQUS using the classical Johnson-Cook constitutive model, the RNO trained in this work and the PCA-Net trained in our previous work  \cite{liu2022learning}, and ensure that the calculation is conducted on a single CPU for proper comparison.  We do a number of  calculations with various time steps (with the same boundary condition imposed all three models), and plot the cost as a function of the number of time steps in Figure \ref{fig:cost}.  We find that the cost scales linearly with the number of time steps in all three models\footnote{We anticipated in our previous work \cite{liu2022learning} that the PCA-Net would scale quadratically since we have to provide the full strain history.  However, this is incorrect: because of the way we
encode memory globally through the PCA basis, this particular history-dependent
calculation scales quadratically with the PCA dimension and linearly with the number of time steps.}. The RNO has a computational cost that is comparable to the classical Johnson-Cook constitutive model.  Further, the PCA-Net is slower than the other two due to the increased cost of communicating the entire strain history as opposed to the strain increment and internal variables.  In summary, the RNO is computationally competitive with traditional models.

\section{Discussion} \label{sec:disc}

In this paper, we have proposed exploiting a recurrent neural operator (RNO) of the form (\ref{eq:RNO}) as a surrogate for a fine-scale model in multi-scale modeling of history dependent materials.   The architecture of the RNO builds in causality in that the stress at any instant of time depends only on the prior history of the deformation.  The history is represented using a { internal}  variable and thus the stress can be evaluated locally in space and time.    The architecture of the RNO is consistent with common formulation of constitutive relations using internal variables or state variables in continuum mechanics \cite{rice1971inelastic,gurtin_cont_mech}.  We may regard the { internal variables to be those} variables that describe the state of the continuum, (\ref{eq:RNO})$_1$ as the stress-strain relation for a given state of the continuum, and (\ref{eq:RNO})$_2$ as the kinetic relation that describes evolution of the state of the continuum.   However, a crucial point is that we do not specify {\it a priori} what the { internal} variables are, but seek to learn them from the stress-strain data.

We have demonstrated in the setting of elasto-viscoplasticity in a polycrystalline material that the RNO can be trained to provide an accurate representation of a polycrystalline ensemble (Section \ref{sec:learn}), and that the trained RNO can be efficiently used in a macroscopic application scale calculation (Section \ref{sec:multi}).  Thus, the trained RNO effectively provides multiscale, specifically FE$^2$, accuracy at a cost comparable to a conventional empirical constitutive relation.

In a previous work, we had explored the use of a PCA-Net as a surrogate constitutive model at the macroscale \cite{liu2022learning}.  The PCA-Net does not have a causal architecture\footnote{A trained PCA-Net, however, does learn causality from the data.} and seeks to approximate the response over the entire map.   The proposed RNO requires significantly less data to train and has a higher accuracy in both the two and three dimensional settings.   Further the current RNO can be evaluated locally while the PCA-Net can not.  For this and other reasons,  it has a computational cost that is smaller than the PCA-Net. 
Finally, we are restricted to computation in a fixed time-interval on which the PCA is conducted; in constrast the RNO can be used on arbitrarily long time intervals as long as the stretch trajectory remains within the realm of the training data.  Therefore, the RNO is a better surrogate in such history dependent materials.  We anticipate that the RNO will similarly outperform other architectures like the graph neural operators \cite{li_multipole_2020} and Fourier neural operator \cite{li_fourier_2021} in this context.

In this work, we have explored multi-scale modeling in the context of elasto-viscoplasticity with crystal plasticity as the fine-scale model.  However, the framework and the methodology is broader and can be used in other settings where we have (numerical or experimental) 
information about the map from the history of deformation to the stress.  Therefore, we can use it in other two-scale settings like  atomic-continuum transitions and granular materials.  We have studied viscoelastic composite materials in a related work \cite{viscoelasticity}.  Further, it can also be used to identify constitutive behavior from experimental data provided we have a sufficiently rich data-set concerning the map from the history of deformation to the stress.

In formulating this work, we recognize that the macroscopic behavior can be an operator from one function space (deformation gradient over an interval of time) to a tensor (though there may be examples where this is also a function space).   This means that we want the trained model to be independent of the resolution (time step) at which the training data is presented and the trained model is used.  In other words, the training data can be at one resolution, but we can use the trained model accurately at every resolution.  This is crucial for the use of this model in multiscale modeling since we may generate training data from the microscopic model at one (generally fine) resolution, but may want to use the trained model in macroscopic calculations at different (generally coarse) resolution.  We have demonstrated resolution-independence of the RNO architecture (\ref{eq:RNO}) in elasto-viscoplasticity in this work.

We believe that the proper architecture, one that correctly reflects the physics is critical for this resolution independence.  For example, while the architecture (\ref{eq:RNO}) shows resolution independence in elasto-viscoplasticity, the slightly different (more general) viscoelastic architecture (\ref{eq:RNO2}) that explicitly includes the stretch rate is not resolution independent when trained with the same data.  This indicates that the effective behavior does not have an explicit strain dependence.  We demonstrated the exact opposite in our work on viscoelasticity \cite{viscoelasticity} where we have explicit strain dependence: the rate independent form (\ref{eq:RNO}) is not while the rate dependent form (\ref{eq:RNO2}) is.

An important, and interesting question, concerns the number of { state} or internal variables that one should use.   We show through theoretical considerations in Section \ref{sec:1d} that we need a single internal variable for elasto-viscoplastic composite (\ref{eq:1dfield}) in one dimension irrespective of the number of constituents.  This is also borne out in the training: we trained RNOs with various numbers of internal variables and found that one is the smallest number that gives rise to good approximation.  {  Further, the discovered state variables in this case happens to be a good approximation of the theoretical state variables\footnote{One can not expect this is general due to reparametrization invariance, but happens to be case here.}.  This gives us confidence in the robustness of the architecture and the training procedure.}

{  In the case of elasto-viscoplastic laminates, Section \ref{sec:lam}, we find empirically that we need three (four with hardening) internal variables for a good approximation while the theory suggests six (respectively eight).  We therefore studied the state variables in the micromechanical model and learnt that the governing equations forced the six (respectively eight) model variables to be confined close to a three (respectively four) dimensional manifold.
}

In the case of elasto-viscoplastic polycrystals, we find empirically through our training procedure that the number of internal variables is related to the dimension of the isochoric manifold -- two in two dimensions, and five in three dimensions.  This is somewhat unexpected.  A typical empirical elasto-viscoplastic law in three dimensions has five variables for the plastic strain, one for the isotropic hardening and five for the kinematic hardening.  The fact that we are able to fit complex loading paths, exhibiting both isotropic and kinematic hardening and using only five variables, suggests that either there is some redundancy in classical empirical models or that there are particular simplifying aspects of the crystal plasticity single crystal model that we use to generate the data.  

The single crystal model we use to generate our data has two particular choices that may be relevant.  First, it uses an isotropic ``compressible neo-Hookean'' elastic model.  This means that the polycrystal is elastically homogeneous.
Second, the various slip and twin systems all have a uniform rate-hardening exponent.  Since different systems are activated in different grains depending on orientation, different rate-hardening exponent can give rise to different time scales which interact in such a manner that requires additional internal variables.  In related recent work on viscoelasticity \cite{viscoelasticity}, we study the overall behavior of a one dimensional composite made of Kelvin-Voigt materials.  There, it is known that a composite with $N$ constituents (with $N$ distinct ratios of elastic modulus to viscosity) requires $N-1$ internal variables.  Further, it is known that in higher dimensions, viscoelastic materials may need an infinite number of internal variables in the presence of corners \cite{brenner2013overall}.  Thus, it is possible that the choices we make in the single crystal model lead to this low number of { internal} variables.  A more complete understanding of this issue remains a topic for the future.  

We note the { internal} variables that we identify do not necessarily have any inherent meaning.  Indeed, as noted in Section \ref{sec:arch}, a change of variables (\ref{eq:cov}) leaves the form of the RNO invariant.  

{ Another interesting question concerns thermodynamic and symmetry restrictions.  In this work, we have have not required any thermodynamic or material symmetry restrictions on the functions $f, g_i$ in (\ref{eq:RNO}).  Instead, we expect the models to learn such restrictions from the data.  It is possible to build architectures with such conditions (e.g., \cite{masi_2021} for dissipative processes and \cite{as'ad_2022} for hyperelasticity).  We do not do so for two reasons.   Thermodynamic and symmetry conditions often restrict the constitutive relations to certain (nonlinear) manifolds, and this breaks the linear space structure that is useful for neural networks.  Further, we believe the ability to learn such non-trivial conditions from data are a very useful check on the ability of these otherwise black-box over-parametrized models to approximate the data.  So there is an interesting balance to be achieved between building in all known physics on the one hand, and maintaining a relevant mathematical structure to learn physics from the data on the other.   This is a topic of current research.

Yet another topic of current research is the connection between the field variables in the micromechanical calculations and the internal variables learned from the macroscopic response.  Here, we train our RNO using only macroscopic stress-strain data following the framework of homogenization or multiscale modeling where the large scale provides the control and the small scale returns the average.  However, our micromechanical simulations have a lot more information, and this information is discarded.  The question of whether this information can be used is a very interesting question, one that is an active topic of our research.  In the example of laminates, Section 4,  we find that the six (eight with strain hardening) dimensional micromechanical (field) variables can be encoded using only three (respectively four) internal variables.  This is very much in the spirit of reduced order modeling.   Further these encoding variables can be used as internal variables.  So it is possible that adding the micromechanical fields can lead to an identification of internal variables with less data (fewer trajectories).   The converse (reconstructing the micromechanical field from the macroscopically learned internal variable) is subtler.  Since the macroscale response is averaged, it is possible that many different micromechanical fields give rise to the same incremental macroscale response.  In other words, the map from micromechanical field to incremental macroscopic response is unique but the converse may not be.  The analogous question over a trajectory over long periods of time is not clear since the flow may collapse the micromechanical fields.  All of this is a topic of current research.}

Finally, we note that our approach does not address the starting microstructure, and that we have to train an RNO for each starting microstructure.  The hidden variables do incorporate some information about the microstructure during its evolution.  Therefore, it is possible that one can incorporate starting microstructure into this RNO framework.  This remains a topic for the future.

{
\renewcommand\thesection{Appendix \Alph{section}}
\setcounter{section}{0}
\renewcommand\theequation{A\arabic{equation}}
\setcounter{equation}{0}
\section{Crystal plasticity in 2D} \label{sec:app}

The deformation gradient is decomposed multiplicatively into the elastic and plastic parts $F=F^\text{e}F^\text{p}$.  We have $N_s$ slip systems with slip plane normal $\{b_i\}_{i=1}^{N_s}$, slip direction $\{b_i\}_{i=1}^{N_s}$, slip activity $\gamma = \{\gamma_i\}_{i=1}^{N_s}$ and accumulated activity $e = \{e_i\}_{i=1}^{N_s}$.  The rate of change plastic deformation gradient and accumulated slip activity depend on the rate of change of slip activity $\dot{F}^\text{p} = \sum_{i=1}^{N_s} \dot{\gamma}_i b_i \otimes n_i$, $\dot{e}_i = |\dot{\gamma}_i|$.

The stored energy density of the polycrystal consists of an elastic energy density $W^\text{e}$ that depends on the elastic part of the deformation gradient $F^\text{e}$ and a stored energy density of plastic work $W^\text{p}$ that depends on the accumulated plastic activity in the slip systems $e$. We take (for 2 dimensions)
\begin{eqnarray}
W^\text{e}(A) &=& \frac{\mu}{2} \left( \frac{\text{tr } A^T A}{\det A} - 2 \right) + \frac{K}{2}(\det A - 1)^2 \label{eq:elas}\\
W^\text{p} &=& \sum_{i = 1}^{N_s} \sigma^\infty_i \left( e_i + \frac{\sigma^\infty_i}{h_i}\exp\left(-\frac{h_i e_i}{\sigma^\infty_i}\right) \right) + \frac{1}{2} \sum e_i  {\mathcal K}_{ij} e_j
\end{eqnarray}
where the first term in the stored energy of plastic work is associated with self-hardening and the second the latent hardening.  Above, $\mu,K$ are the shear and bulk moduli; $\sigma^\infty_i$ is the ultimate strength and $h_i$ the hardening parameter on the $i^\text{th}$ slip system; and $K_{ij}$ is the latent hardening matrix with zero diagonal entries.  The evolution of the slip activity is governed by the dissipation potential
\begin{eqnarray}
    \Psi^*(\gamma) = \sum_{i = 1}^{N_s} \frac{\tau^{\, \text{p}}_{0, \, i}\dot{\gamma}^{\, \text{p}}_{0, \, i}}{m_{i} + 1} \bigg(\frac{|\dot{\gamma^{\, \text{p}}}_i|}{\dot{\gamma}^{\, \text{p}}_{0, \, i}}\bigg)^{m_{i} + 1}
\end{eqnarray}
where  $\tau^{\, \text{p}}_{0, \, i}$ is the critical resolved shear stresses, $\dot{\gamma}^{\, \text{p}}_{0, \, i}$ the reference shear rate, and $m_{i}$ the power rate hardening parameter for the $i^\text{th}$ slip system.

The system evolves satisfying mechanical equilibrium and the yield criterion,
\begin{equation}
\nabla \cdot \left(\frac{\partial W}{\partial F}\right) = 0 \quad \text{and} \quad
0 \in - \tau^{\, \text{p}}_i + \frac{\partial W_{\, \text{p}}}{\partial \epsilon^{\, \text{p}}_i} + \partial_{\dot{\epsilon}^{\, \text{p}}_i} \Psi^{*} 
\end{equation}
respectively, along with the boundary conditions at each time.
}

\section*{Acknowledgement} 

 We gratefully acknowledge the financial support of the U.S. Army Research Laboratory through cooperative agreement W911NF-12-2-0022 (BL, AS, KB) and award W911NF22-2-0120 (AS, KB), the U.S. Army Research Office through award W911NF-22-1-0269 (KB).
AS is also supported as a Department of Defense Vannevar Bush Faculty Fellow.  MT is funded by the Department of Energy Computational Science Graduate Fellowship under Award Number DE-SC002111.

The views and conclusions contained in this document are those of the authors and should not be interpreted as representing the official policies, either expressed or implied, of the Army Research Laboratory or the U.S. Government. The U.S. Government is authorized to reproduce and distribute reprints for Government purposes notwithstanding any copyright notation herein.


\end{document}